\documentclass[fleqn]{article}

\usepackage{graphicx}
\usepackage{amsmath}

\def\tref#1#2#3{{\bf #1} (#2) #3}
\def\bref#1{(\ref{#1})}
\def\unt#1{\,{\rm #1}}

\begin{document}

\title{
{\sf 
Gaussian Sum-Rules and Prediction of Resonance Properties
} }

\author{
G.\ Orlandini,
T.G.\ Steele 
\thanks{Permanent Address: Department of Physics \& Engineering Physics, University of Saskatchewan,
Saskatoon, SK ~~S7N 5E2, Canada}
\\
{\sl 
Dipartimento di Fisica and}\\
{\sl 
INFN Gruppo Collegato di Trento}\\
{\sl Universit\`a di Trento}\\
{\sl I-38050 Povo, Italy}
\\[10pt]
D.\ Harnett\\
{\sl Department of Physics \& Engineering Physics}\\
{\sl University of Saskatchewan}\\
{\sl Saskatoon, Saskatchewan S7N 5E2, Canada.}
}
\maketitle
\begin{abstract}
Techniques for using Gaussian QCD sum-rules to predict hadronic resonance properties are developed 
for single-resonance and two-resonance phenomenological models, and criteria are developed for determining which
of these models is required for analyzing a particular hadronic channel.  The vector current sum-rule coupled to the
$\rho$ meson  is shown to be consistent with a single resonance model, and the Gaussian sum-rule analysis results in an 
accurate $\rho$ mass prediction which exhibits excellent agreement 
between the theoretical prediction of the Gaussian sum-rule and the phenomenological model.   
A two-resonance model is shown to be necessary for the Gaussian sum-rule for the non-strange quark scalar ($\bar n n$) 
currents.  The
two-resonance Gaussian sum-rule analysis of the isoscalar and isovector ($I=0,1$) $\bar n n$ scalar mesons  
  exhibits excellent agreement between the theoretical prediction and phenomenological model.  The prediction of the
resonance properties of the $I=0,1$   $\bar n n$ scalar mesons in this two-resonance model provides valuable 
information for the interpretation of the scalar mesons, including the $X(1775)$.
\end{abstract}

\section{Introduction}\label{intro_sec}
QCD  sum-rules are an established technique for relating hadronic properties to theoretical 
QCD predictions.  The most frequently used sum-rules for this purpose are the Laplace \cite{SVZ} and
finite-energy sum-rules \cite{fesr}.   The significance  of using the finite-energy sum-rule (FESR)  as a 
supplementary constraint on a Laplace sum-rule analysis was most clearly established through the 
Gaussian sum-rules \cite{gauss}
\begin{equation}
G\left( \hat s,\tau\right)=\frac{1}{\pi}\int\limits_{t_0}^\infty\frac{1}{\sqrt{4\pi\tau}}
\exp{\left( -\frac{\left(t-\hat s\right)^2}{4\tau}\right)}\,\rho(t)\,dt
\quad ,\quad \tau>0
\label{basic_gauss}
\end{equation}
where $\rho(t)$ is a hadronic spectral function with  physical threshold $t_0$.  
The quantity $G\left( \hat s,\tau\right)$ on the left-hand side of 
\bref{basic_gauss} is determined from a theoretical calculation of a correlation function 
$\Pi\left(Q^2\right)$.  The  composite operators used to construct the correlation function serve as 
interpolating fields for the hadronic channel represented by $\rho(t)$ on the right-hand side of
\bref{basic_gauss}.

The Gaussian kernel in \bref{basic_gauss}   implies that $G\left( \hat s,\tau\right)$ satisfies a diffusion 
equation \cite{gauss}.
\begin{equation}
\frac{\partial^2 G\left( \hat s,\tau\right)}{\partial\hat s^2}=
\frac{\partial G\left( \hat s,\tau\right)}{\partial \tau}
\label{diffusion}
\end{equation}
The relation between Gaussian and finite-energy sum-rules was established \cite{gauss} by showing that the resonance plus continuum model for $\rho(t)$, when evolved through the diffusion equation, would only 
reproduce the QCD prediction at large energies (large $\tau$) if the resonance and continuum threshold $s_0$
were related by the  $n=0$ member of the following FESR family.
\begin{equation}
F_n\left(s_0\right)=\frac{1}{\pi}\int\limits_{t_0}^{s_0} t^n\rho(t)\,dt \quad ,\quad n=0,1,2,\ldots
\label{basic_fesr}
\end{equation}
As with the Gaussian sum-rule \bref{basic_gauss}, the left-hand side of \bref{basic_fesr} is obtained
from a QCD prediction.

Sum-rules with Gaussian-like (Lorentzian)  kernels have already proven to have  strong predictive power in nuclear physics
\cite{nuclear}.
The purpose of this paper is to demonstrate that Gaussian sum-rules can also be used predictively in 
hadronic physics,  and to develop techniques for the analysis of Gaussian sum-rules in hadronic physics.  
Section \ref{concept_sec}
reviews the formulation and conceptual advantages of Gaussian sum-rules, and develops techniques for analyzing 
a single-resonance plus continuum phenomenological model.  Section \ref{rho_sec} applies these 
techniques to the $\rho$ meson  which has 
traditionally been used to establish the validity of sum-rule techniques.  
Finally in Section 
\ref{scalar_sec} we study the phenomenologically challenging case of the quark scalar mesons probed through 
the non-strange $\bar n n$ current.  The scalar sector requires an extension of the analysis techniques to a 
two-resonance phenomenological model, and development of criteria to determine when such an extension is necessary.
This sensitivity of the Gaussian sum-rule to excited states is one of the unique features of the Gaussian sum-rule  
compared with  Laplace sum-rules.  The phenomenological implications of the results of the Gaussian sum-rule analysis of 
the $\bar n n$ quark scalar mesons will be presented at the end of Section \ref{scalar_sec}.

\section{Conceptual Foundations of Gaussian Sum-Rules}\label{concept_sec}
Consider the basic form of the Gaussian sum-rule \bref{basic_gauss}.   Leaving aside the method for 
calculating 
$G\left( \hat s,\tau\right)$,  consider the $\tau\to 0$ limit of the sum-rule.  Using the identity
\begin{equation}
\lim_{\tau\to 0}\frac{1}{\sqrt{4\pi\tau}}
\exp{\left( -\frac{\left(t-\hat s\right)^2}{4\tau}\right)}=\delta\left(t-\hat s\right)
\label{delta_fn}
\end{equation}  
we see that
\begin{equation}
\lim_{\tau\to 0}G\left(\hat s,\tau\right)=\frac{1}{\pi}\rho\left(\hat s\right)\quad ,\quad \hat s>t_0
\label{gauss_limit}
\end{equation}
Since $\rho(t)$ is related to  hadronic physics quantities, the $\tau\to 0$ limit of the Gaussian 
sum-rule would in principle allow direct extraction of hadronic physics from the $\hat s$ dependence of 
$G\left(\hat s,\tau\right)$ obtained from QCD.

Although the above limit cannot be achieved in practice, the property \bref{gauss_limit} demonstrates an
important conceptual advantage of the Gaussian sum-rule.  Consider the equivalent form of the 
Laplace sum-rule
\begin{equation}
R\left(\Delta^2\right)=\frac{1}{\pi}\int\limits_{t_0}^\infty\exp{\left(-\frac{t}{\Delta^2}\right)}\rho(t)\, dt
\quad .
\label{basic_laplace}
\end{equation}
The $\Delta\to 0$ limit (again not possible to achieve in practice) would in principle only emphasize the
$t=0$ region of the integration region, and would not reveal the energy dependence of 
$\rho(t)$.

Now consider a more realistic case of a Gaussian sum-rule with $\tau>0$.  The Gaussian kernel
is peaked at $t=\hat s$, and has a width of $\sqrt{2\tau}$.  Thus $G\left(\hat s,\tau\right)$ represents a 
smearing of $\rho(t)$ in the region $\hat s-\sqrt{2\tau}< t<\hat s+\sqrt{2\tau}$, centered at
$t=\hat s$.    It is therefore reasonable to hope that with $\tau$ fixed at a reasonable physical value, 
the $\hat s$ dependence of $G\left(\hat s ,\tau\right)$  will reveal the rough structure of 
$\rho(t)$.
For example, if $t=m_r^2$ corresponds to a sharp resonance peak of $\rho(t)$, 
then one would also expect a peak in $G\left(\hat s,\tau\right)$ near $\hat s=m_r^2$.  These expectations 
are  upheld by the analysis of the $\rho$ and scalar quark mesons in subsequent sections.  

The concept of quark-hadron duality, where averaged hadronic quantities are equivalent to the QCD 
prediction,
is explicitly manifested in the parameter $\tau$ which controls the size of the region near $t=\hat s$ over 
which  hadronic physics is averaged.  As will be discussed below, $\tau$ is directly related to the energy 
scale for running quantities in the renormalization-group improvement of the theoretical prediction 
$G\left(\hat s,\tau\right)$, and no constraint on the parameter $\hat s$ emerges.  The limitations of 
QCD originating from the renormalization scale then provide a natural duality interval for
reasonable agreement between hadronic physics and QCD.  Finally, we shall see that nonperturbative 
contributions to $G\left(\hat s ,\tau\right)$ are suppressed with increasing $\hat s$, explicitly 
reinforcing the importance of non-perturbative effects in the low-energy region.

The above aspects of the relation between QCD and hadronic physics 
 illuminated by  Gaussian sum-rules are obscured  in the Laplace sum-rules.  The reason for this contrast
is that the Laplace kernel $\exp{\left(-t/\Delta^2\right)}$ exponentially suppresses  the entire 
energy region of 
$\rho(t)$, while the Gaussian kernel only damps the energy region of $\rho(t)$ away from 
$t=\hat s$.  Thus the two scales $\hat s$ and $\tau$ in the Gaussian sum-rule provide a more detailed 
probe of hadronic physics than  the Laplace sum-rule.

\subsection{Single-Resonance Analysis of Gaussian Sum-Rules}
\label{sing_res_sec}
Extraction of hadronic properties from the theoretical prediction $G\left(\hat s,\tau\right)$ requires
a phenomenological model for $\rho(t)$.  
The ``resonance plus continuum''  model is used to 
represent the hadronic physics phenomenology  contained in the integral of $\rho(t)$.  
In this model,  
 hadronic physics is (locally) dual to the theoretical QCD prediction
for energies above the continuum threshold $t=s_0$
\begin{equation}
\rho(t)=\theta\left(s_0-t\right)\rho^{had}(t)+\theta\left(t-s_0\right){\rm Im}\Pi^{QCD}(t)
\label{res_plus_cont}
\end{equation} 
where $s_0$ is the continuum
threshold above which hadronic physics and QCD are locally dual.
The continuum contribution to the integral
of \bref{basic_gauss} is denoted by
\begin{equation}
G^{cont}\left(\hat s,\tau, s_0\right)=
\frac{1}{\pi}\int\limits_{s_0}^\infty\frac{1}{\sqrt{4\pi\tau}}
\exp{\left( -\frac{\left(t-\hat s\right)^2}{4\tau}\right)}\,{\rm Im}\Pi^{QCD}(t)\,dt
\label{continuum}
\end{equation}
Since the continuum contribution is determined by QCD, it is combined with the theoretical quantity
$G\left(\hat s,\tau\right)$ 
\begin{equation}
G^{QCD}\left(\hat s, \tau,s_0\right)\equiv
G\left(\hat s, \tau\right)-G^{cont}\left(\hat s, \tau,s_0\right)
\label{gauss_QCD}
\end{equation}
resulting in a Gaussian sum-rule relating QCD to hadronic physics phenomenology.
\begin{equation}
G^{QCD}\left( \hat s,\tau,s_0\right)=\frac{1}{\pi}\int\limits_{t_0}^{s_0}\frac{1}{\sqrt{4\pi\tau}}
\exp{\left( -\frac{\left(t-\hat s\right)^2}{4\tau}\right)}\,\rho^{had}(t)\,dt
\label{QCD_had}
\end{equation}

Consider the  single narrow resonance 
model for $\rho^{had}(t)$ widely (but not exclusively) employed in the analysis of Laplace sum-rules
\begin{equation}
\frac{1}{\pi}\rho^{had}(t)=f_r^2\delta\left(t-m_r^2\right)
\label{nar_res}
\end{equation}
where $m_r$ denotes the resonance mass, and $f_r^2$ is the integrated resonance strength.
Use of this model in \bref{QCD_had}
leads to a Gaussian sum-rule which relates QCD to the properties of the single resonance.
\begin{equation}
G^{QCD}\left(\hat s, \tau,s_0\right)=\frac{f_r^2}{\sqrt{4\pi\tau}}\exp{\left[-
\frac{\left(\hat s-m_r^2\right)^2}{4\tau}\right]}
\label{gauss_1res}
\end{equation}

The continuum threshold $s_0$ is constrained by the finite-energy sum
rule \bref{basic_fesr}, which in the single  narrow resonance model becomes
\begin{equation}
F_0\left(s_0\right)=f_r^2\quad .
\label{fesr_1nr}
\end{equation}
Since the finite-energy sum-rule must be satisfied for the Gaussian sum-rule to evolve asymptotically to 
the QCD prediction through the diffusion equation \bref{diffusion} (heat-evolution test \cite{gauss}), the
normalization of the sum-rule \bref{gauss_QCD} is fixed by the finite-energy sum-rule constraint, as
can be verified by taking the $\hat s$ integral of \bref{gauss_1res}.
\begin{equation}
\int\limits_{-\infty}^\infty G^{QCD}\left(\hat s,\tau, s_0\right)d\hat s=f_r^2
\label{gauss_norm}
\end{equation}
Thus since the quantity $f_r^2$ is constrained by the finite-energy sum-rule \bref{fesr_1nr},
the normalization of the quantity $G^{QCD}\left(\hat s,\tau, s_0\right)$ 
is already determined.  Consequently,  the information in the sum-rule \bref{gauss_1res} remaining after imposing the 
heat-evolution test \cite{gauss} is contained in the quantity $N^{QCD}\left(\hat s, \tau, s_0\right)$ normalized to
unit area
\begin{gather}
N^{QCD}\left(\hat s, \tau, s_0\right)= 
\frac{G^{QCD}\left(\hat s, \tau, s_0\right)}{M_0\left(\tau, s_0\right)}
\label{norm_sr}\\
M_0\left( \tau, s_0\right)=\int\limits_{-\infty}^\infty G^{QCD}\left(\hat s,\tau, s_0\right)d\hat s
\quad ,
\label{m0}
\end{gather}  
leading to a sum-rule independent of the resonance-strength and finite-energy sum-rule.
\begin{equation}
N^{QCD}\left(\hat s, \tau, s_0\right)= \frac{1}{\sqrt{4\pi\tau}}\exp{\left[-
\frac{\left(\hat s-m_r^2\right)^2}{4\tau}\right]}
\label{norm_gauss}
\end{equation}
 
Viewed as a function of $\hat s$, the phenomenological (right-hand) side of \bref{norm_gauss} has a 
peak (maximum) which occurs at $\hat s=m_r^2$ independent of $\tau$, and the peak height  $1/\sqrt{4\pi\tau}$
at this maximum 
 shows $\tau$ dependence. These properties will be used to find the optimum value of 
$s_0$ from which $m_r^2$ can be extracted.  This optimum value of $s_0$ can then be used to determine the 
total resonance strength through the finite-energy sum-rule \bref{fesr_1nr}.

The analysis proceeds by fixing $\tau$, and finding the value $\hat s_{peak}\left(\tau,s_0\right)$ at which 
$N^{QCD}\left(\hat s,\tau,s_0\right)$ has a maximum\footnote{Our numerical analysis uses the Golden search algorithm 
\cite{num_recipes}  to determine $\hat s_{peak}\left(\tau,s_0\right)$}
\begin{equation}
\left.\frac{d}{d\hat s}N^{QCD}\left(\hat s,\tau,s_0\right)
\right|_{\hat s=\hat s_{peak}\left(\tau,s_0\right)}=0
\label{s_peak}
\end{equation}
and then determining the peak height 
\begin{equation}
N_{peak}\left(\tau,s_0\right)=N^{QCD}\left(\hat s_{peak}\left(\tau,s_0\right),\tau,s_0\right)\quad .
\label{N_peak}
\end{equation}
The optimum value of $s_0$ is constrained by the requirements following from the properties of the
phenomenological side of \bref{norm_gauss}:
\begin{enumerate}
\item $\hat s_{peak}\left(\tau,s_0\right)=m_r^2$ is constant ($\tau$-independent),
\item $\sqrt{4\pi\tau}\,N_{peak}(\tau,s_0)=1$ (independent of $\tau$).
\end{enumerate}
Thus the optimum value of $s_0$, and the corresponding prediction of the resonance mass $m_r^2$, can 
be determined by minimizing a $\chi^2$ measure related to the above properties
\begin{equation}
\chi^2\left(s_0\right)=\sum_{i=1}^n\left(
\frac{\hat s_{peak}\left(\tau_i,s_0\right)}{m_r^2}-1\right)^2+
\sum_{i=1}^n\left(
N_{peak}\left(\tau_i,s_0\right)\sqrt{4\pi\tau_i}-1
\right)^2
\label{chi2}
\end{equation}
where $\tau_i$ are equally-spaced points in the $\tau$ region of interest, and 
 $m_r^2$ is implicitly a function of $s_0$
\begin{equation}
m_r^2=\frac{1}{n}\sum_{i=1}^n\hat s_{peak}\left(\tau_i,s_0\right)\quad .
\label{1nr_predict}
\end{equation}
The value of $s_0$ which minimizes the $\chi^2$ thus provides the best agreement with properties of the
phenomenological model and leads to a prediction of the resonance mass $m_r^2$ in the single 
narrow-resonance model.

Before proceeding with the application of these $\chi^2$ techniques to the $\rho$ meson, extensions of 
the narrow resonance model will be discussed to determine the magnitude of effects that could arise from 
resonance widths $\Gamma$ and resonance shapes.  Intuitively, one would expect that if the Gaussian width 
is much wider than the resonance width, then the width effects will be negligible.  

The simplest extension of the unit-area narrow resonance $\delta\left(t-m_r^2\right)$ is the unit area 
square pulse previously used for studying width effects in the Laplace sum-rules \cite{square}
\begin{equation}
\frac{1}{\pi}\rho^{sp}(t)=
\frac{1}{2m_r\Gamma}\left[
\theta\left(t-m^2_R+m_r\Gamma\right)-\theta\left(t-m^2_R-m_r\Gamma\right)
\right]\quad ,
\label{square_pulse}
\end{equation}
which has the following Gaussian image
\begin{equation}
\begin{split}
G^{sp}\left(\hat s, \tau\right)& =\frac{1}{2m_r\Gamma\sqrt{4\pi\tau}}
\int\limits_{m_r^2-m_r\Gamma}^{m_r^2+m_r\Gamma}
\exp{\left( -\frac{\left(t-\hat s\right)^2}{4\tau}\right)}\,dt
\\
&=\frac{1}{4m_r\Gamma}\left[
{\rm erf}\left(\frac{\hat s-m^2_R+m_r\Gamma}{2\sqrt{\tau}}\right)
-{\rm erf}\left(\frac{\hat s-m^2_R-m_r\Gamma}{2\sqrt{\tau}}
\right)
\right]\quad ,
\end{split}
\label{gauss_sp}
\end{equation}
where
\begin{equation}
{\rm erf}(x)=\frac{2}{\sqrt{\pi}}\int\limits_0^x e^{-y^2}\,dy\quad .
\label{erf}
\end{equation}
Such a resonance model clearly overestimates the effect of resonance widths compared with a Breit-Wigner 
which is more concentrated about the resonance peak.
The expression \bref{gauss_sp} still has a peak position at $\hat s=m_r^2$ as in the narrow resonance model, but its 
peak height is 
altered to 
\begin{equation}
G^{sp}\left(\hat s=m_r^2, \tau\right)=\frac{1}{2m_r\Gamma}{\rm erf}\left(
\frac{m_r\Gamma}{2\sqrt{\tau}}
\right)
\approx\frac{1}{\sqrt{4\pi\tau}}\left[1
-\frac{m_r^2\Gamma^2}{12\tau}+{\cal O}\left(\Gamma^4\right)
\right]
\label{gauss_sp_limit}
\end{equation}
As anticipated, the resonance width effects in \bref{gauss_sp_limit} diminish with 
increasing Gaussian width $\tau$.  
Using the full expression in terms of the error function,  the deviation from the 
narrow width limit is less than 5\% for $m_r\Gamma/({2\sqrt{\tau}})<0.4$, a condition which is 
easily satisfied for even a wide resonance such as the $f_0(1370)$ in the $\tau>1\unt{GeV^4}$ range.
Such uncertainties are well below those associated with the theoretical prediction 
of $G\left(\hat s,\tau\right)$.  
For the $\rho$ meson, the  resonance 
width  is a completely negligible 0.2\% effect even for $\tau=0.5\unt{GeV^4}$.
Thus we conclude that for reasonable ranges of the the Gaussian width $\tau$, the Gaussian sum-rule is   
insensitive to the effect of resonance widths, and so the
narrow width phenomenological model is in fact  an accurate description of non-zero width resonances
in the Gaussian sum-rules.

\subsection{Formulation of Gaussian Sum-Rules}
In this section we briefly summarize salient features of the formulation of Gaussian sum-rules  
presented in \cite{gauss}.  Consider a dispersion relation with one subtraction constant as needed for
the correlation function of vector currents used to probe the $\rho$ meson\footnote{Extension to correlation functions
requiring further subtraction constants leads to a final result identical to (\protect\ref{borel_gauss}) obtained for a single subtraction constant.}.
\begin{equation}
\Pi\left(Q^2\right)=\Pi(0)-\frac{Q^2}{\pi}\int\limits_{t_0}^\infty
dt\,\frac{ \rho(t)}{t\left(t+Q^2\right)}
\label{dispersion}
\end{equation}
The difference between the dispersion relation at $Q^2=-\hat s+i\Delta$ and $Q^2=-\hat s-i\Delta$  
cancels the dependence on the subtraction constant $\Pi(0)$.
\begin{equation}
\frac{\Pi\left(-\hat s-i\Delta\right)-\Pi\left(-\hat s+i\Delta\right)}{2i\Delta}=
\frac{1}{\pi}\int\limits_{t_0}^\infty
dt\,\frac{ \rho(t)}{\left(t-\hat s\right)^2+\Delta^2}
\label{gauss_1}
\end{equation}
For large $\Delta$ the left-hand side of \bref{gauss_1} is determined by QCD since 
it evaluates the correlation function well away from the physical cut.  
As with Laplace sum-rules \cite{SVZ}, the Gaussian sum-rule can be  obtained from the dispersion relation 
\bref{gauss_1}
through  the Borel transform operator
 $\hat B$
\begin{equation}
\hat B\equiv 
\lim_{\stackrel{N \rightarrow \infty~,~\Delta^2\rightarrow \infty}{\Delta^2/N\equiv 4\tau}}
\frac{\left(-\Delta^2\right)^N}{\Gamma(N)}\left(\frac{d}{d\Delta^2}\right)^N
\label{borel}
\end{equation} 
which has the following property relevant to construction of the Gaussian sum-rule
\begin{equation}
\hat B \left[ \frac{1}{x+\Delta^2}\right]=\frac{1}{4\tau} \exp{\left(-\frac{x}{4\tau}\right)}  \quad .
\label{borel_exp}
\end{equation}
Defining the theoretically-determined quantity
\begin{equation}
G\left(\hat s,\tau\right)=\sqrt{\frac{\tau}{\pi}}\hat B\left[
\frac{\Pi\left(-\hat s-i\Delta\right)-\Pi\left(-\hat s+i\Delta\right)}{i\Delta}
\right]
\label{borel_gauss}
\end{equation}
leads to the Gaussian sum-rule \bref{basic_gauss} after application of $\hat B$ to \bref{gauss_1}.

\begin{figure}[htb]
\centering
\includegraphics[scale=0.5,angle=270]{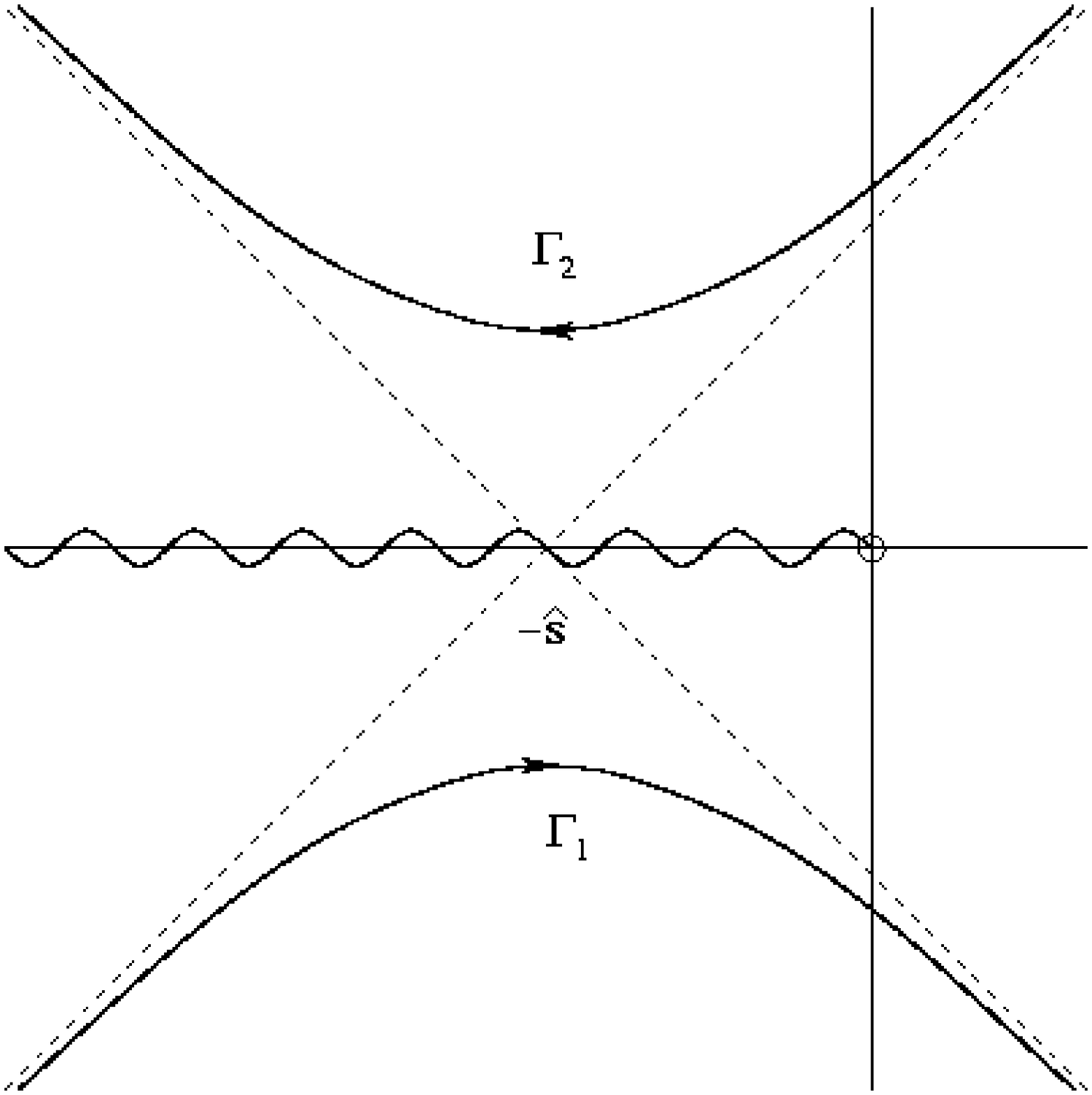}
\caption{Contour of integration $\Gamma_1+\Gamma_2$ defining the 
Gaussian  sum-rule  in (\protect\ref{cont_gauss_w}). 
The wavy line 
on the negative real axis denotes the branch cut of $\Pi(w)$.
}
\label{cont_fig}
\end{figure}

An alternative to the direct calculation of the Gaussian sum-rules through the definition
of $\hat B$ in \bref{borel} is obtained through an identity relating the Borel and
Laplace transform \cite{gauss}.  
\begin{gather}
f\left(\Delta^2\right)=\int\limits_0^\infty F(T) e^{-\Delta^2 T}\, dT\equiv{\cal L}\left[ F(T)\right]
\quad \Longrightarrow\quad \frac{1}{T}\hat B\left[ f\left(\Delta^2\right)\right]
=F(T)={\cal L}^{-1}
\left[ f\left(\Delta^2\right)\right]\quad ,\quad T\equiv \frac{1}{4\tau}
\label{borel_laplace}\\
{\cal L}^{-1}
\left[ f\left(\Delta^2\right)\right]=\frac{1}{2\pi i}\int\limits_{b-i\infty}^{b+i\infty}
f\left(\Delta^2\right) e^{\Delta^2 T}\,d\Delta^2
\label{inv_lap_def}
\end{gather}
where the real parameter $b$ in the definition \bref{inv_lap_def} of the inverse Laplace 
transform must be chosen so that $f\left(\Delta^2\right)$ is analytic to the right of the 
contour of integration in the complex $\Delta^2$ plane. Using the result \bref{borel_laplace}, the Gaussian sum-rule
\bref{borel_gauss} can be obtained from an inverse Laplace transform
\begin{equation}
\begin{split}
G\left(\hat s,\tau\right)&=\frac{1}{4\sqrt{\pi\tau}}{\cal L}^{-1}\left[
\frac{\Pi\left(-\hat s-i\Delta\right)-\Pi\left(-\hat s+i\Delta\right)}{i\Delta}
\right]
\\
&=\frac{1}{4\sqrt{\pi\tau}}\frac{1}{2\pi i}\int\limits_{b-i\infty}^{b+i\infty}
\frac{\Pi\left(-\hat s-i\Delta\right)-
\Pi\left(-\hat s+i\Delta\right)}{i\Delta}
\exp{\left(\Delta^2 T\right) \,d\Delta^2} \quad .
\end{split}
\label{cont_gauss1}
\end{equation}
Respectively mapping the individual terms in \bref{cont_gauss1} containing  $\Pi\left(-\hat s \pm i\Delta\right)$ 
from the $\Delta^2$ to the $w=-\hat s \pm i\Delta$ 
complex plane results in the following expression for the Gaussian sum-rule
\begin{equation}
G\left(\hat s,\tau\right)=\frac{1}{\sqrt{4\pi\tau}}\frac{1}{2\pi i}\int\limits_{\Gamma_1+\Gamma_2}
\!\!\Pi(w)\exp{\left[-\frac{\left(w+\hat s\right)^2}{4\tau}\right]}\, dw
\label{cont_gauss_w}
\end{equation} 
where the contours $\Gamma_1$ and $\Gamma_2$ are illustrated in Figure \ref{cont_fig}.

\section{Prediction of the $\rho$ Mass from Gaussian Sum-Rules}\label{rho_sec}
The $\rho$ meson is probed through the vector-isovector current correlation function.
\begin{gather}
\Pi_{v}\left(Q^2\right)\left[q^\mu q^\nu-q^2g^{\mu\nu}\right]=
i\int d^4x\,e^{iq\cdot x}\left\langle O \right.\left\vert T\left[ J^\mu(x) J^\nu(0)\right]\right\vert 
\left. O\right\rangle
\quad ,\quad Q^2=-q^2>0
\label{rho_corr_fn}\\
J^\mu(x)=\frac{1}{2}\left[ \bar u(x)\gamma^\mu u(x)-\bar d(x)\gamma^\mu d(x)\right]
\label{vec_current}
\end{gather}
The field-theoretical (QCD) calculation of $\Pi_v\left(Q^2\right)$ consists of perturbative (logarithmic)
corrections and QCD vacuum effects of infinite correlation length parametrized by the power-law 
corrections from the QCD vacuum condensates \cite{SVZ}
\begin{equation}
\Pi_v\left(Q^2\right)=\Pi^{pert}_v\left(Q^2\right)+\Pi^{cond}_v\left(Q^2\right) \quad .
\end{equation}
To two-loop order in the  chiral limit $m_u=m_d=0$
the perturbative contribution\footnote{  
 A divergent constant has been ignored since like the subtraction constant $\Pi(0)$, it 
vanishes  in the formation of the Gaussian sum rule.} in the  
$\overline{\rm MS}$ scheme for three active flavours is \cite{vec_pert}
\begin{equation}
\Pi^{pert}_v\left(Q^2\right)=-\frac{1}{8\pi^2}\log{\left(\frac{Q^2}{\nu^2}\right)}
\left[1+\frac{\alpha(\nu)}{\pi}\right]
\label{vec_pert}
\end{equation}
where $\nu$ is the renormalization scale.  To lowest order, the QCD condensate contributions 
up to dimension 8 are given by \cite{SVZ}
\begin{gather}
\Pi^{cond}_v\left(Q^2\right)=\frac{1}{8\pi^2}\left[ 
\frac{1}{Q^4}\left\langle C_4^v{\cal O}^v_4\right\rangle+
\frac{1}{Q^6}\left\langle C_6^v{\cal O}^v_6\right\rangle+
\frac{1}{Q^8}\left\langle C_8^v{\cal O}^v_8\right\rangle
\right]
\label{vec_cond}\\
\left\langle C^v_4{\cal O}^v_4\right\rangle=\frac{\pi}{3}\left\langle \alpha G^2\right\rangle
-8\pi^2 m\left\langle \bar q q\right\rangle
\label{c4_vec}\\
\left\langle C^v_6{\cal O}^v_6\right\rangle=-\frac{896}{81}\pi^3\alpha\left(\left\langle
\bar q q\right\rangle\right)^2
\label{c6_vec}
\end{gather}
where  $SU(2)$ symmetry and the vacuum saturation hypothesis have been employed.  For brevity, we refer 
to the literature \cite{dim_eight} for the 
expressions 
for the dimension eight condensates, and simply use \bref{vec_cond} to establish a normalization 
consistent with \cite{bordes}.

\begin{figure}[htb]
\centering
\includegraphics[scale=0.5,angle=270]{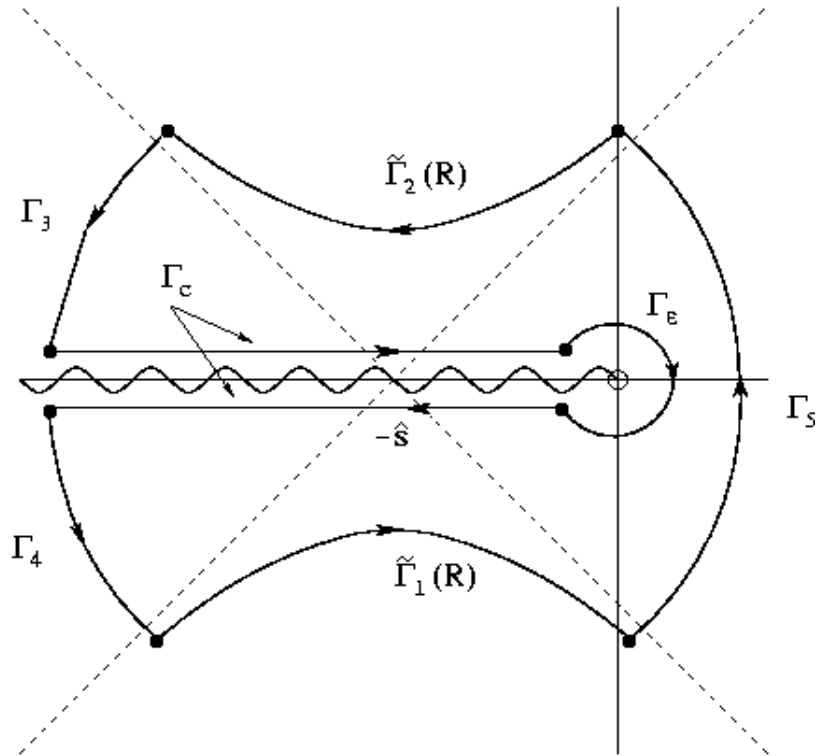}
\caption{Closed contour $C(R)$ used to 
calculate the Gaussian sum-rule defined by  (\protect\ref{cont_gauss_w}). 
The  inner circular segment $\Gamma_\epsilon$ has a radius of $\epsilon$, and the  circular 
segments 
$\Gamma_3$, $\Gamma_4$ and $\Gamma_5$
have a radius $R$.
The wavy line 
on the negative real axis denotes the branch cut of $\Pi(w)$, 
and the linear segments of the contour above and below the branch cut are denoted by $\Gamma_c$.
The contour $\tilde{\Gamma}_1(R)$ is that portion of $\Gamma_1$ (see Figure \protect\ref{cont_fig})
which lies in the interior of a circle of radius $R$ centred at $-\hat s$, and similarly
for  $\tilde{\Gamma}_2(R)$.
}
\label{cont_fig2}
\end{figure}

To evaluate the  Gaussian sum-rule consider the contour $C(R)$ in Figure \ref{cont_fig2}. The quantity $\Pi(w)$ 
is analytic within and on $C(R)$; consequently
\begin{equation}
0=\frac{1}{\sqrt{4\pi\tau}}\frac{1}{2\pi i}\oint\limits_{C(R)}\!\!\Pi(w)
\exp{\left[-\frac{\left(w+\hat s\right)^2}{4\tau}\right]}\, dw
\quad .
\label{residue}
\end{equation}
In the limit as $R\to\infty$, the integrals along $\Gamma_3$, $\Gamma_4$ and $\Gamma_5$ all approach 
zero for the QCD expressions given in (\ref{vec_pert},~\ref{vec_cond}).   Furthermore $\tilde{\Gamma}_1(R)$ and 
$\tilde{\Gamma}_1(R)$ respectively approach $\Gamma_1$ and $\Gamma_2$ as $R\to\infty$.  We therefore obtain
the following expression  for the Gaussian sum-rule \bref{cont_gauss_w}
\begin{equation}
\begin{split}
G\left(\hat s,\tau\right)=&-\frac{1}{\sqrt{4\pi\tau}}\frac{1}{2\pi i}
\lim_{R\to\infty}
\int\limits_{\Gamma_c+\Gamma_\epsilon}
\!\!\Pi(w)\exp{\left[-\frac{\left(w+\hat s\right)^2}{4\tau}\right]}\, dw
 \\
=&-\frac{1}{\sqrt{4\pi\tau}}\frac{1}{2\pi i}\int\limits_\epsilon^\infty
\left[\Pi\left(te^{i\pi}\right) -\Pi\left(te^{-i\pi}\right)\right]
\exp{\left(-\frac{\left(t-\hat s\right)^2}{4\tau}\right)}\,dt
\\ 
&+\frac{1}{\sqrt{4\pi\tau}}\frac{1}{2\pi}\int\limits_{-\pi}^\pi
\epsilon e^{i\theta}\Pi\left(\epsilon e^{i\theta}\right)
\exp{\left(-\frac{\left(\epsilon e^{i\theta}+\hat s\right)^2}{4\tau}\right)} \,d\theta\quad .
\end{split}
\label{def_cont_gauss1}
\end{equation} 

The perturbative contributions from the  $\Gamma_\epsilon$ contour ($\theta$ integral) in \bref{def_cont_gauss1}
will be zero in the $\epsilon\to 0$ limit, leaving only the integral of the discontinuity across the 
branch cut [{\it i.e.} ${\rm Im}\Pi^{pert}(t)$] to determine the perturbative contributions to the 
Gaussian sum-rule
\begin{equation}
G^{pert}\left(\hat s, \tau\right)=
\frac{1}{\sqrt{4\pi\tau}}\int\limits_0^\infty\frac{1}{8\pi^2} \left(1+\frac{\alpha(\nu)}{\pi}\right)
\exp{\left(-\frac{\left(t-\hat s\right)^2}{4\tau}\right)}\,dt
=\frac{1}{16\pi^2} \left(1+\frac{\alpha(\nu)}{\pi}\right)
\left[1+{\rm erf}\left(\frac{\hat s}{2\sqrt{\tau}}\right)\right] \quad ,
\label{g_pert_vec}
\end{equation}
Since the current \bref{vec_current} is renormalization-group invariant, ${\rm Im}\Pi^{pert}(t)$ 
satisfies a homogeneous renormalization-group equation.  The result of renormalization-group improvement 
for the Gaussian sum-rules can  be inferred from  the general structure of perturbative 
corrections which take the form
\begin{equation}
\int\limits_0^\infty \log^n\left(\frac{t}{\nu^2}\right) 
\exp{\left(-\frac{\left(t-\hat s\right)^2}{4\tau}\right)}\,dt=
\nu^2\int\limits_0^\infty dx\,\log^n(x)\exp{\left[-\left(
\frac{\nu^2x}{2\sqrt{\tau}}-\frac{\hat s}{2\sqrt{\tau}}
\right)^2\right]}=\nu^2 H\left(\frac{\nu^2}{\sqrt{\tau}}, \frac{\hat s}{\sqrt{\tau}}\right)\quad .
\label{rg_scale}
\end{equation}
The functional dependence of the perturbative corrections to the Gaussian sum-rules expressed by the function $H$ 
demonstrates that $\nu$ scales with $\sqrt{\tau}$.  Thus, the solution of the 
renormalization-group  equation satisfied by the perturbative  contributions is the replacement of  
$\alpha(\nu)$ with  the running coupling constant and identifies the scale $\nu^2=\sqrt{\tau}$.
This implies the existence of  low-energy boundary (lower bound) on $\tau$, but places no restriction 
on the scale $\hat s$.  
 
The QCD condensate contributions  \bref{vec_cond} to $\Pi\left(Q^2\right)$ do not have a branch 
discontinuity, so their contribution to the Gaussian sum-rule arises solely from the contour
$\Gamma_\epsilon$ ($\theta$ integral) in \bref{def_cont_gauss1}, and can be evaluated using the result
\begin{equation}
-\frac{1}{2\pi i}\int\limits_{\Gamma_\epsilon}\!\!
\frac{1}{w^n}\exp{\left[-\frac{\left(w+\hat s\right)^2}{4\tau}\right]}\, dw
=\lim_{w\to 0}\frac{1}{(n-1)!}\frac{d^{n-1}}{dw^{n-1}}
\exp{\left[-\frac{\left(w+\hat s\right)^2}{4\tau}\right]}
\quad , \quad n=1,2,3\ldots\quad .
\label{g_cond}
\end{equation}

Expressions   \bref{g_pert_vec}, \bref{g_cond}, \bref{vec_cond}, and \bref{def_cont_gauss1}
lead to the Gaussian sum-rule of vector currents.  
\begin{equation}
\begin{split}
G\left(\hat s, \tau\right)
=&\frac{1}{16\pi^2} \left(1+\frac{\alpha\left(\sqrt{\tau}\right)}{\pi}\right)
\left[1+{\rm erf}\left(\frac{\hat s}{2\sqrt{\tau}}\right)\right]
-\frac{\hat s}{32\pi^2\tau\sqrt{\pi\tau}}\exp{\left(-\frac{\hat s^2}{4\tau}\right)}
\left\langle C^v_4{\cal O}^v_4\right\rangle
\\
 &+\frac{1}{64\pi^2\tau\sqrt{\pi\tau}}\left(-1+\frac{\hat s^2}{2\tau}\right)
\exp{\left(-\frac{\hat s^2}{4\tau}\right)}
\left\langle C^v_6{\cal O}^v_6\right\rangle
-\frac{\hat s}{128\pi^2\tau^2\sqrt{\pi\tau}}\left(-1+\frac{\hat s^2}{6\tau}\right)
\exp{\left(-\frac{\hat s^2}{4\tau}\right)}
\left\langle C_8^v{\cal O}^v_8\right\rangle
\end{split}
\label{gauss_vec}
\end{equation}
Agreement between  \bref{gauss_vec} and \cite{gauss} provides a useful check on the conventions 
established in  \bref{cont_gauss_w}.
We also note that the QCD condensate contributions to \bref{gauss_vec} can be recovered by
expanding the Gaussian kernel in \bref{basic_gauss}
about $t=0$ (or alternatively in  a series about any other value of $t$, or in a 
series of Hermite polynomials), and then using the relation between the QCD condensates and  the FESR family \bref{basic_fesr}
(in addition, see equation (6.15) 
in reference \cite{gauss}). This demonstrates that in principle 
({\it e.g.} knowledge of  higher-dimension condensate contributions),  
sum-rules based on the Gaussian (or Laplace) kernel
contain the same information  as the whole FESR family.  
However, as we discussed previously and will be demonstrated below, the   Gaussian kernel  
arranges this information more effectively, 
enhancing the  predictive power compared with other sum-rule kernels.

The QCD continuum contributions \bref{continuum} arising from the perturbative corrections to the vector 
correlator are
\begin{equation}
G^{cont}\left(\hat s, \tau,s_0\right)
=\frac{1}{16\pi^2} \left(1+\frac{\alpha\left(\sqrt{\tau}\right)}{\pi}\right)
\left[1-{\rm erf}\left(\frac{s_0-\hat s}{2\sqrt{\tau}}\right)\right]
\label{vec_continuum}
\end{equation}
and hence the theoretically determined Gaussian sum-rule quantity \bref{gauss_QCD} 
for the vector currents is
\begin{equation}
\begin{split}
G^{QCD}\left(\hat s, \tau, s_0\right)
=&\frac{1}{16\pi^2} \left(1+\frac{\alpha\left(\sqrt{\tau}\right)}{\pi}\right)
\left[{\rm erf}\left(\frac{\hat s}{2\sqrt{\tau}}\right)
+{\rm erf}\left(\frac{s_0-\hat s}{2\sqrt{\tau}}\right)
\right]
-\frac{\hat s}{32\pi^2\tau\sqrt{\pi\tau}}\exp{\left(-\frac{\hat s^2}{4\tau}\right)}
\left\langle C^v_4{\cal O}^v_4\right\rangle
\\
&+\frac{1}{64\pi^2\tau\sqrt{\pi\tau}}\left(-1+\frac{\hat s^2}{2\tau}\right)
\exp{\left(-\frac{\hat s^2}{4\tau}\right)}
\left\langle C_6^v{\cal O}^v_6\right\rangle
\\
&-\frac{\hat s}{128\pi^2\tau^2\sqrt{\pi\tau}}\left(-1+\frac{\hat s^2}{6\tau}\right)
\exp{\left(-\frac{\hat s^2}{4\tau}\right)}
\left\langle C_8^v{\cal O}^v_8\right\rangle\quad .
\end{split}
\label{gauss_qcd_vec}
\end{equation}
Finally, the quantity $M_0\left(\tau, s_0\right)$ in \bref{m0}  required for the 
normalized vector-current Gaussian sum-rule
\bref{norm_sr}  is given by
\begin{equation}
M_0\left( \tau, s_0\right)=\int\limits_{-\infty}^\infty G^{QCD}\left(\hat s,\tau, s_0\right)d\hat s
=\frac{1}{8\pi^2} \left(1+\frac{\alpha\left(\sqrt{\tau}\right)}{\pi}\right)
s_0\quad .
\label{m0_vec}
\end{equation}

The non-perturbative QCD condensate contributions in \bref{gauss_qcd_vec} are exponentially suppressed for large
$\hat s$.  Since $\hat s$ represents the location of the Gaussian peak on the phenomenological side of the sum-rule,
the non-perturbative corrections are most important in the low-energy region, as anticipated by the 
role of  QCD condensates in relation to the vacuum properties of QCD.  This explicit low-energy role of the QCD
condensates clearly exhibited for the Gaussian sum-rules is obscured in the Laplace sum-rule because the peak of the 
Laplace exponential kernel in \bref{basic_laplace}
is always located at $t=0$.  

Several QCD parameters must be specified before carrying out the phenomenological analysis of the
Gaussian sum-rule of vector currents.  The running coupling for three active flavours to 
two-loop order is
\begin{equation}
\frac{\alpha_s\left(\nu^2\right)}{\pi}= \frac{1}{\beta_0 L}-\frac{\beta_1\log L}{\beta_0\left(\beta_0L\right)^2}
\quad ,\quad L=\log\left(\frac{\nu^2}{\Lambda^2}\right)\quad ,\quad
\beta_0=\frac{9}{4}\quad ,\quad \beta_1=4
\label{alpha}
\end{equation}
with 
 $\Lambda_{\overline{MS}}\approx 300\,{\rm MeV}$ for three active flavours,
consistent with current estimates of $\alpha_s(M_\tau)$ \cite{PDG,pade_tau} and matching conditions 
through the charm threshold \cite{ChetyrkinKniehl}.
For the gluon condensate we use the central value  determined in \cite{bordes}
\begin{equation}
\left\langle \alpha G^2\right\rangle=\left(0.045\pm 0.014\right)\unt{GeV^4}\quad ,
\label{aGG_cond}
\end{equation}
and PCAC \cite{GMOR} is used for the quark condensate.
\begin{equation}
m\left\langle \bar q q\right\rangle=\frac{1}{2}m_u\langle \bar u u\rangle+\frac{1}{2}m_d\langle \bar d d\rangle
=-\frac{1}{2}f_\pi^2m_\pi^2\quad ,\quad f_\pi=93\unt{MeV}\quad .
\label{qq_cond}
\end{equation}
Deviations of the dimension-six condensate from the  vacuum saturation value are  parameterized by 
the quantity $f_{vs}$  which could be as large as $f_{vs}=2$ \cite{bordes,dim_six}
\begin{equation}
\left\langle C_6^v{\cal O}^v_6\right\rangle=-f_{vs}\frac{896}{81}\pi^3\alpha\left(\left\langle
\bar q q\right\rangle\right)^2=-f_{vs}\frac{896}{81}\pi^3\left(1.8\times 10^{-4}\unt{GeV^6}\right)
\quad ,\quad f_{vs}=1.5\pm 0.5\quad .
\label{qqq_cond}
\end{equation}
Finally, we use the reference \cite{bordes} value for the dimension-eight condensate.
\begin{equation}
\left\langle C_8^v{\cal O}^v_8\right\rangle=\left(0.40\pm 0.16\right)\unt{GeV^8}
\label{dim_eight}
\end{equation}

The $\chi^2$ minimization techniques outlined in Section \ref{sing_res_sec} [see equations \bref{chi2} and 
\bref{1nr_predict}] are now  employed to determine the optimum continuum threshold  $s_0$ and 
the corresponding prediction of the $\rho$ meson mass $m_\rho$, 
demonstrating the  ability of Gaussian sum-rules to predict resonance properties.   
Using  equally-spaced points in the range\footnote{Altering the range to 
$1.0\unt{GeV^4}\le\tau\le 4.0\unt{GeV^4}$ has  minimal effect on the predictions.}
$0.5\unt{GeV^4}\le\tau\le 4.0\unt{GeV^4}$ for determining 
 $\chi^2\left(s_0\right)$ in \bref{chi2}, and 
with inclusion of the 
uncertainties of the  QCD condensates given in (\ref{aGG_cond}--\ref{dim_eight}) we obtain the
following results for $m_\rho$ and $s_0$
\begin{equation}
m_\rho=(0.75\pm 0.07)\unt{GeV}\quad ,\quad s_0=(1.2\pm 0.2)\unt{GeV^2}
\label{rho_predict}
\end{equation}
in excellent agreement with the measured mass $m_\rho=770\unt{MeV}$. 

A detailed examination of the required equality  \bref{norm_gauss} between the Gaussian  sum-rule 
and the phenomenological single resonance model 
can be obtained by using the $\chi^2$-predicted values of $m_\rho$ and $s_0$
as input into the single-resonance plus continuum model, and
examining the $\hat s$, $\tau$ dependence of 
the sum-rule in comparison with the phenomenological model. 
Figure \ref{rho_fig} compares this $\hat s$ dependence of the 
sum-rule  and phenomenological model for selected $\tau$ values in the region used to define
the $\chi^2$.  As would be anticipated by its construction, the $\chi^2$ optimization procedure
leads to excellent correspondence between theory and phenomenology for the $\hat s_{peak}$ position and 
peak height.  However, the astounding agreement between the $\hat s$ dependence of the sum-rule and 
phenomenological model provides strong evidence for the ability of Gaussian sum-rules to predict hadronic 
properties. In the next section, Gaussian sum-rules will be employed to study the (non-strange) $\bar n n$ 
scalar mesons,  a more challenging channel for hadronic physics phenomenology.

\section{Gaussian Sum-Rule Analysis of the Quark Scalar Mesons}\label{scalar_sec} 
The interpretation  of the scalar mesons is a challenging problem in hadronic physics since  
a variety of interpretations exist for the lowest-lying scalar resonances 
($\sigma$ or $f_0(400-1200)$, $f_0(980)$, $f_0(1370)$, $f_0(1500)$, $a_0(980)$, $a_0(1450)$ \cite{PDG})
including conventional quark-antiquark ($q\bar q$) states, $K\bar K$ molecules, gluonium, four-quark models, 
and dynamically generated thresholds \cite{interp,Speth,Jaffe}. 

The relevant correlation function for the $I=0,1$ non-strange $\bar n n$ scalar mesons is
\begin{equation}
\Pi_s\left(Q^2\right)= i \,\int\,d^4x\,e^{iq\cdot x} \langle 0 | T \left[J_s (x) J_s (0)\right] |0\rangle
\quad ,\quad Q^2=-q^2>0
\label{correlator}
\end{equation}
where, in the $SU(2)$ limit [$m_q \equiv (m_u + m_d)/2$] for isoscalar $(I = 0)$
and isovector $(I = 1)$ currents,
\begin{equation}
J_s(x) = m_q\left[\overline{u}(x)u(x) + (-1)^I\, \overline{d}(x)d(x)\right]/{2} .
\label{current}
\end{equation}
We note that the factor of the quark mass is necessary for a renormalization-group invariant current.
Correlation functions of 
scalar and pseudoscalar currents  are unique since they receive significant contributions
from instantons \cite{instanton,ins_liquid}.    In contrast to QCD condensates which represent vacuum effects of 
infinite correlation length, instanton contributions represent finite correlation-length 
QCD vacuum effects, and are the only known theoretical mechanism that distinguishes between the $I=0$ and $I=1$ 
correlation functions  in the presence of $SU(2)$ flavour symmetry \cite{isobreaking}.  Thus the theoretical
calculation of the  
scalar current correlation function consists of perturbative, condensate, and instanton contributions
\begin{equation}
\Pi_s\left(Q^2\right)=\Pi^{pert}_s\left(Q^2\right)+\Pi^{cond}_s\left(Q^2\right) 
+\Pi^{inst}_s\left(Q^2\right)\quad .
\label{scalar_parts}
\end{equation}
At two-loop order and to leading order in the quark mass for three active flavours in the $\overline{MS}$ scheme, 
the perturbative contributions in \bref{scalar_parts} are  
\cite{Chetyrkin}\footnote{A field-theoretical divergence proportional to $Q^2$ 
is ignored since it vanishes after application of $\hat B$ in the formation of the Gaussian sum-rule.} 
\begin{equation}
\Pi_s^{pert}\left(Q^2\right)=\frac{3 m_q^2 Q^2}{16\pi^2}\log{\left(\frac{Q^2}{\nu^2}\right)}
\left[1+\frac{\alpha}{\pi}\left(-\log{\left(\frac{Q^2}{\nu^2}\right)} +\frac{17}{3}\right)\right]\quad .
\label{scalar_pert}
\end{equation}
To leading order in $m_q$ and $\alpha$, the QCD condensate contributions in \bref{scalar_parts} are
\cite{Bagan,Reinders,SVZ}:
\begin{gather}
\Pi_s^{cond}(Q^2) = m_q^2\left[\frac{\left\langle C_4^s {\cal O}_4^s\right\rangle}{Q^2} 
+\frac{\left\langle C_6^s {\cal O}_6^s\right\rangle}{Q^4}\right] 
\label{condensates}\\
\left\langle C_4^s {\cal O}_4^s\right\rangle=
\frac{3}{2} \left\langle m_q \overline{q}q\right\rangle +
\frac{1}{16\pi} \left\langle\alpha_s G^2 \right\rangle
\label{c4_scalar}
\\
\langle C^s_6{\cal O}^s_6\rangle= \pi\alpha_s
\left[ 
\frac{1}{4}\left\langle \left(\bar u\sigma_{\mu\nu}\lambda^a u-\bar d\sigma_{\mu\nu}\lambda^a d\right)^2\right\rangle
+\frac{1}{6}
\left\langle \left(   
\bar u \gamma_\mu \lambda^a u+\bar d \gamma_\mu \lambda^a d 
\right)
\sum_{u,d,s}\bar q \gamma^\mu \lambda^a q
\right\rangle\right]\quad .
\label{o6}
\end{gather}
As for the vector current correlation function, the  vacuum saturation hypothesis \cite{SVZ} in the $SU(2)$ limit
$\langle \bar u u\rangle=\langle\bar d d\rangle\equiv\langle\bar q q\rangle$ provides  a reference value
for $\langle {\cal O}^s_6\rangle$
\begin{equation}
\left\langle C_6^s{\cal O}^s_6\right\rangle=-f_{vs}\frac{88}{27}\alpha_s
\left\langle (\bar q q)^2\right\rangle
=-f_{vs}1.8\times 10^{-4} {\rm GeV}^6\quad .
\label{o61}
\end{equation}
As mentioned earlier, the perturbative and condensate contributions to the scalar correlation function are
independent of $I$, and hence
do not distinguish between the isoscalar and isovector channels.

Using  symmetry properties relating the instanton contributions to 
pseudoscalar and scalar correlation functions \cite{isobreaking} combined with the 
instanton liquid model results for the pseudoscalar correlation function \cite{SVZ,ins_liquid}, 
the instanton contributions to the scalar correlation function are
\begin{equation}
\Pi^{inst}_s\left(Q^2\right)=\left(-1\right)^I\frac{3m_q^2Q^2}{4\pi^2}\left[
K_1\left(\rho_c\sqrt{Q^2}\right)\right]^2
\quad ,
\label{pi_inst}
\end{equation}
where $K_1$ is a modified Bessel function \cite{abramowitz}, and $\rho_c=1/\left(600\unt{MeV}\right)$ is the
(uniform) instanton size in the instanton liquid model \cite{ins_liquid}. The explicit factor 
depending on $I$ in \bref{pi_inst} is the only theoretical contribution that distinguishes between the isovector and
isoscalar channels.

Calculation of the Gaussian sum-rule for the scalar currents proceeds as in Section \ref{rho_sec}.  Using 
the following asymptotic property  of the modified Bessel function \cite{abramowitz}
\begin{equation}
K_1\left(z\right)\sim\sqrt{\frac{\pi}{2z}}e^{-z}\quad ;\quad |z|\gg 1~,~\left|{\rm arg}(z)\right|\le\frac{3\pi}{2}
\end{equation} 
one can verify that the QCD expressions given in (\ref{scalar_pert},~\ref{condensates},~\ref{pi_inst})
uphold the 
transition from \bref{residue} to \bref{def_cont_gauss1} developed for the vector currents.  Thus we can calculate
the Gaussian sum-rule for scalar currents from \bref{cont_gauss1}.

As with the vector channel, the perturbative contributions \bref{scalar_pert} from the contour 
 $\Gamma_\epsilon$ ($\theta$ integral) in \bref{def_cont_gauss1}
will be zero in the $\epsilon\to 0$ limit, leaving only the integral of the discontinuity across the 
branch cut [{\it i.e.} ${\rm Im}\Pi^{pert}(t)$] to determine the perturbative contributions to the 
Gaussian sum-rule.
\begin{equation}
G^{pert}\left(\hat s, \tau\right)=
\frac{1}{\sqrt{4\pi\tau}}\frac{3m_q^2\left(\sqrt{\tau}\right)}{16\pi^2}
\int\limits_0^\infty 
\left[t\left(1+\frac{17}{3}\frac{\alpha\left(\sqrt{\tau}\right)}{\pi}\right)
-2\frac{\alpha\left(\sqrt{\tau}\right)}{\pi}t\log{\left(\frac{t}{\sqrt{\tau}}\right)}
\right]
\exp{\left(-\frac{\left(t-\hat s\right)^2}{4\tau}\right)}\,dt
\label{g_pert_scalar}
\end{equation}
This expression has already been renormalization-group improved 
by the replacement of $\nu^2=\sqrt{\tau}$ and implicit 
identification of $\alpha$ and $m_q$ as running quantities at this same scale.  
The running coupling has already been given
in \bref{alpha}, and the running quark mass for three active flavours at two-loop order is
\begin{equation}
m_q\left(\nu^2\right)=  \frac{\hat m_q}{\left(\frac{1}{2}L\right)^{\frac{4}{9}}}\left(
1+\frac{290}{729}\frac{1}{L}-\frac{256}{729}\frac{\log{ L}}{L}
\right) \quad  ,\quad L=\log\left(\frac{\nu^2}{\Lambda^2}\right)\quad ,
\label{run_mass}
\end{equation}
where $\hat m_q$ is the renormalization-group invariant quark mass parameter.

As for the vector channel, the QCD condensate contributions  \bref{condensates} do not have a branch 
discontinuity, so their contribution to the Gaussian sum-rule arises solely from the contour
$\Gamma_\epsilon$ ($\theta$ integral) in \bref{def_cont_gauss1}, and can be evaluated using \bref{g_cond}.

The instanton contributions to the Gaussian sum-rule follow from \bref{cont_gauss1}
\begin{equation}
\begin{split}
G^{inst}\left(\hat s,\tau\right)
=&
\left(-1\right)^I\frac{3m_q^2}{4\pi^2}\frac{1}{\sqrt{4\pi\tau}}\frac{1}{2\pi i}\int\limits_\epsilon^\infty
t\left(
\left[K_1\left(\rho_c\sqrt{t}e^{i\frac{\pi}{2}}\right)\right]^2
-\left[K_1\left(\rho_c\sqrt{t}e^{-i\frac{\pi}{2}}\right)\right]^2
\right)
\exp{\left(-\frac{\left(t-\hat s\right)^2}{4\tau}\right)}\,dt
\\
& +
\left(-1\right)^I\frac{3m_q^2}{4\pi^2}
\frac{1}{\sqrt{4\pi\tau}}\frac{1}{2\pi}\int\limits_{-\pi}^\pi
\epsilon^2 e^{i2\theta}\left[K_1\left(\rho_c\sqrt{\epsilon} e^{i\frac{\theta}{2}}\right)\right]^2
\exp{\left(-\frac{\left(\epsilon e^{i\theta}+\hat s\right)^2}{4\tau}\right)} \,d\theta\quad .
\end{split}
\label{g_inst1}
\end{equation} 
Simplification of \bref{g_inst1} requires the following properties of the modified
Bessel function $K_1(z)$ \cite{abramowitz} 
\begin{gather}
K_1\left(z\right)\sim\frac{1}{z}\quad ,\quad z\to 0
\label{asymp_K1}\\
K_1\left(z\right)=\left\{
\begin{array}{l}
-\frac{\pi}{2}H_1^{(1)}\left(ze^{i\pi/2}\right)\quad ,\quad -\pi<\arg(z)\le \frac{\pi}{2}\\[5pt]
-\frac{\pi}{2}H_1^{(2)}\left(ze^{-i\pi/2}\right)\quad ,\quad -\frac{\pi}{2}<\arg(z)\le \pi
\end{array}
\right.
\label{anal_con_K1}
\end{gather}
where $H_1^{(1)}(z)=J_1(z)+iY_1(z)$ and $H_1^{(2)}(z)=J_1(z)-iY_1(z)$.
The asymptotic behaviour \bref{asymp_K1} implies that the $\theta$ integral of \bref{g_inst1}
will be zero in the $\epsilon\to 0$ limit  and the identity \bref{anal_con_K1}
allows evaluation of the discontinuity in the $t$ integral of  \bref{g_inst1}, leading to the
following instanton contribution to the Gaussian sum-rule
\begin{equation}
G^{inst}\left(\hat s,\tau\right)=-\left(-1\right)^I\frac{3 m_q^2}{8\pi}
\frac{1}{\sqrt{4\pi\tau}}\int\limits_0^\infty
t J_1\left(\rho_c\sqrt{t}\right) Y_1\left(\rho_c\sqrt{t}\right) 
\exp{\left(-\frac{\left(t-\hat s\right)^2}{4\tau}\right)}\,dt \quad .
\label{g_inst_fin}
\end{equation}

Expressions \bref{g_pert_scalar}, \bref{g_cond}, \bref{condensates}, \bref{def_cont_gauss1} and
\bref{g_inst_fin} lead to the Gaussian sum-rule of scalar currents
\begin{equation}
\begin{split}
G\left(\hat s, \tau\right)=&
\frac{1}{\sqrt{4\pi\tau}}\frac{3m_q^2\left(\sqrt{\tau}\right)}{16\pi^2}
\int\limits_0^\infty 
\left[t\left(1+\frac{17}{3}\frac{\alpha\left(\sqrt{\tau}\right)}{\pi}\right)
-2\frac{\alpha\left(\sqrt{\tau}\right)}{\pi}t\log{\left(\frac{t}{\sqrt{\tau}}\right)}
\right]
\exp{\left(-\frac{\left(t-\hat s\right)^2}{4\tau}\right)}\,dt
\\
&-\left(-1\right)^I\frac{3 m_q^2}{8\pi}
\frac{1}{\sqrt{4\pi\tau}}\int\limits_0^\infty
t J_1\left(\rho_c\sqrt{t}\right) Y_1\left(\rho_c\sqrt{t}\right) 
\exp{\left(-\frac{\left(t-\hat s\right)^2}{4\tau}\right)}\,dt
\\
& +\quad m_q^2\exp{\left(-\frac{\hat s^2}{4\tau}\right)}\left[
\frac{1}{2\sqrt{\pi\tau}}\left\langle C^s_4{\cal O}^s_4\right\rangle-
\frac{\hat s}{4\tau\sqrt{\pi\tau}}\left\langle C^s_6{\cal O}^s_6\right\rangle
\right]\quad .
\end{split}
\label{gauss_scalar}
\end{equation}

Comparison of \bref{g_inst_fin} and \bref{continuum} indicates the existence of an instanton continuum contribution
\cite{instanton_continuum}
\begin{equation}
\frac{1}{\pi} \,Im\Pi^{inst}(t) = -\left(-1\right)^I
\frac{3m_q^2}{8\pi}\, tJ_1 (\rho\sqrt{t})Y_1(\rho\sqrt{t}) \quad .
\label{ImPi_ins}
\end{equation}
Combined with the perturbative continuum devolving from \bref{scalar_pert} we obtain the 
continuum contribution to the Gaussian sum-rule of scalar currents
\begin{equation}
\begin{split}
G^{cont}\left(\hat s, \tau\right)=&
\frac{1}{\sqrt{4\pi\tau}}\frac{3m_q^2\left(\sqrt{\tau}\right)}{16\pi^2}
\int\limits_{s_0}^{\infty}
\left[t\left(1+\frac{17}{3}\frac{\alpha\left(\sqrt{\tau}\right)}{\pi}\right)
-2\frac{\alpha\left(\sqrt{\tau}\right)}{\pi}t\log{\left(\frac{t}{\sqrt{\tau}}\right)}
\right]
\exp{\left(-\frac{\left(t-\hat s\right)^2}{4\tau}\right)}\,dt
\\
&-\left(-1\right)^I\frac{3 m_q^2}{8\pi}
\frac{1}{\sqrt{4\pi\tau}}\int\limits_{s_0}^\infty
t J_1\left(\rho_c\sqrt{t}\right) Y_1\left(\rho_c\sqrt{t}\right) 
\exp{\left(-\frac{\left(t-\hat s\right)^2}{4\tau}\right)}\,dt
\end{split}
\label{gauss_scalar_cont}
\end{equation}
and hence the theoretically determined Gaussian sum-rule \bref{gauss_QCD} for the scalar currents
\begin{equation}
\begin{split}
G^{QCD}\left(\hat s, \tau\right)=&
\frac{1}{\sqrt{4\pi\tau}}\frac{3m_q^2\left(\sqrt{\tau}\right)}{16\pi^2}
\int\limits_0^{s_0} 
\left[t\left(1+\frac{17}{3}\frac{\alpha\left(\sqrt{\tau}\right)}{\pi}\right)
-2\frac{\alpha\left(\sqrt{\tau}\right)}{\pi}t\log{\left(\frac{t}{\sqrt{\tau}}\right)}
\right]
\exp{\left(-\frac{\left(t-\hat s\right)^2}{4\tau}\right)}\,dt
\\
&   -\left(-1\right)^I\frac{3 m_q^2}{8\pi}
\frac{1}{\sqrt{4\pi\tau}}\int\limits_0^{s_0}
t J_1\left(\rho_c\sqrt{t}\right) Y_1\left(\rho_c\sqrt{t}\right) 
\exp{\left(-\frac{\left(t-\hat s\right)^2}{4\tau}\right)}\,dt
\\
& + m_q^2\exp{\left(-\frac{\hat s^2}{4\tau}\right)}\left[
\frac{1}{2\sqrt{\pi\tau}}\left\langle C^s_4{\cal O}^s_4\right\rangle-
\frac{\hat s}{4\tau\sqrt{\pi\tau}}\left\langle C^s_6{\cal O}^s_6\right\rangle
\right]\quad .
\end{split}
\label{gauss_scalar_QCD}
\end{equation}
Finally, the quantity $M_0\left(\tau, s_0\right)$ in \bref{m0}  required for the 
normalized scalar-current Gaussian sum-rule
\bref{norm_sr}  is given by
\begin{equation}
\begin{split}
M_0\left( \tau, s_0\right)
=&\frac{3m_q^2\left(\sqrt{\tau}\right)}{16\pi^2}
\left(1+\frac{17}{3}\frac{\alpha\left(\sqrt{\tau}\right)}{\pi}\right)\frac{s_0^2}{2}
-\frac{3m_q^2\left(\sqrt{\tau}\right)}{16\pi^2}\frac{\alpha\left(\sqrt{\tau}\right)}{\pi}\frac{s_0^2}{2}
\left(2\log{\left(\frac{s_0}{\sqrt{\tau}}\right)}-1\right)
\\
&- \left(-1\right)^I\frac{3 m_q^2}{8\pi}
\int\limits_0^{s_0}t J_1\left(\rho_c\sqrt{t}\right) Y_1\left(\rho_c\sqrt{t}\right) \,dt
+m_q^2\left\langle C^s_4{\cal O}^s_4\right\rangle
\end{split}
\label{m0_scalar}
\end{equation}

All theoretical contributions in \bref{gauss_scalar_QCD} and \bref{m0_scalar} are proportional to $m_q^2$, and
hence proportional to the $\hat m_q^2$ through \bref{run_mass}.  Consequently, the normalized Gaussian sum-rule 
$N^{QCD}$ will be {\em independent} of $\hat m_q^2$ which is clearly advantageous given the uncertainty in 
determinations of the light-quark masses \cite{PDG}.

If the Gaussian sum-rules for scalar currents are analyzed using the single-resonance $\chi^2$-optimization 
techniques of Section \ref{sing_res_sec} we find
poor agreement between the theoretical prediction and
the single resonance phenomenological model for the optimized values of the mass and continuum.  
Figures \ref{f0_sing_res_fig} and  \ref{a0_sing_res_fig} show that the peak height of the single resonance model is 
consistently {\em larger} than the theoretical prediction,
and that the peak position of the theoretical prediction 
shows some $\tau$ dependence.  \footnote{We note that  peak drift is absent in the equation
(\protect\ref{gauss_sp}) expression for width effects, and furthermore 
unrealistically large resonance widths would be required to accommodate the observed  discrepancy at the peak height.} 
Furthermore a discrepancy is seen at the tails of the distributions  
where the  single-resonance model becomes {\em smaller} than the theoretical prediction.  This latter point 
indicates that the width of the theoretical distribution is larger than that of the single resonance model.

A quantitative measure of this behaviour
can be found in higher-order moments  of the Gaussian sum-rule.
\begin{gather}
M_k\left(\tau, s_0\right)=\int\limits_{-\infty}^\infty \hat s^k G^{QCD}\left(\hat s,\tau, s_0\right)d\hat s
\quad ,\quad k=0,1,2,3\ldots
\label{mk_gauss}\\
\frac{M_k\left(\tau, s_0\right)}{M_0\left(\tau, s_0\right)}=
\int\limits_{-\infty}^\infty \hat s^k N^{QCD}\left(\hat s,\tau, s_0\right)d\hat s
\quad ,\quad k=0,1,2,3\ldots\quad .
\label{mk_norm}
\end{gather}
Of particular importance are the moments related to the quantities $\sigma^2$ and $A_2$ 
defining the width and asymmetry of the distributions
\begin{gather}
\sigma^2=\frac{M_2}{M_0}-\left(\frac{M_1}{M_0}\right)^2
\label{dist_width}\\
A_2=\frac{M_3}{M_0}-3\frac{M_2}{M_0}\frac{M_1}{M_0}
+2\left(\frac{M_1}{M_0}\right)^3 \quad .
\label{dist_asymm}
\end{gather}
For the scalar channel, the relevant higher-order moments for our analysis are given by
\begin{gather}
\begin{split}
M_1\left( \tau, s_0\right)
=&\frac{3m_q^2\left(\sqrt{\tau}\right)}{16\pi^2}
\left(1+\frac{17}{3}\frac{\alpha\left(\sqrt{\tau}\right)}{\pi}\right)\frac{s_0^3}{3}
-\frac{3m_q^2\left(\sqrt{\tau}\right)}{16\pi^2}\frac{\alpha\left(\sqrt{\tau}\right)}{\pi}\frac{2s_0^3}{9}
\left(3\log{\left(\frac{s_0}{\sqrt{\tau}}\right)}-1\right)
\\
&- \left(-1\right)^I\frac{3 m_q^2}{8\pi}
\int\limits_0^{s_0}t^2 J_1\left(\rho_c\sqrt{t}\right) Y_1\left(\rho_c\sqrt{t}\right) \,dt
-m_q^2\left\langle C^s_6{\cal O}^s_6\right\rangle
\end{split}
\label{m1_scalar}\\
\begin{split}
M_2\left( \tau, s_0\right)
=&\frac{3m_q^2\left(\sqrt{\tau}\right)}{16\pi^2}
\left(1+\frac{17}{3}\frac{\alpha\left(\sqrt{\tau}\right)}{\pi}\right)\left[\frac{s_0^4}{4}+\tau s_0^2\right]
\\
&-\frac{3m_q^2\left(\sqrt{\tau}\right)}{16\pi^2}\frac{\alpha\left(\sqrt{\tau}\right)}{\pi}\frac{s_0^2}{8}
\left[
s_0^2\left(4\log{\left(\frac{s_0}{\sqrt{\tau}}\right)}-1\right)
+8\tau \left(2\log{\left(\frac{s_0}{\sqrt{\tau}}\right)}-1\right)
\right]
\\
&- \left(-1\right)^I\frac{3 m_q^2}{8\pi}
\int\limits_0^{s_0} t\left(t^2+2\tau\right) J_1\left(\rho_c\sqrt{t}\right) Y_1\left(\rho_c\sqrt{t}\right) \,dt
+m_q^2 2\tau\left\langle C^s_4{\cal O}^s_4\right\rangle
\end{split}
\label{m2_scalar}
\\
\begin{split}
M_3\left( \tau, s_0\right)
=&\frac{3m_q^2\left(\sqrt{\tau}\right)}{16\pi^2}
\left(1+\frac{17}{3}\frac{\alpha\left(\sqrt{\tau}\right)}{\pi}\right)\left[\frac{s_0^5}{5}+2\tau s_0^3\right]
\\
&-\frac{3m_q^2\left(\sqrt{\tau}\right)}{16\pi^2}\frac{\alpha\left(\sqrt{\tau}\right)}{\pi}2s_0^3
\left[
\frac{s_0^2}{25}\left(5\log{\left(\frac{s_0}{\sqrt{\tau}}\right)}-1\right)
+\frac{2}{3}\tau \left(3\log{\left(\frac{s_0}{\sqrt{\tau}}\right)}-1\right)
\right]
\\
&- \left(-1\right)^I\frac{3 m_q^2}{8\pi}
\int\limits_0^{s_0} t^2\left(t^2+6\tau\right) J_1\left(\rho_c\sqrt{t}\right) Y_1\left(\rho_c\sqrt{t}\right) \,dt
-m_q^2 6\tau\left\langle C^s_6{\cal O}^s_6\right\rangle
\end{split}
\label{m3_scalar}
\end{gather}
For completeness, we also give the moments needed to calculate $\sigma^2$ for the vector channel.
\begin{gather}
M_1\left( \tau, s_0\right)=\frac{1}{8\pi^2} \left(1+\frac{\alpha\left(\sqrt{\tau}\right)}{\pi}\right)\frac{s^2_0}{2}
-\frac{1}{8\pi^2}\left\langle C_4^v{\cal O}_4^v\right\rangle
\label{m1_vec}\\
M_2\left( \tau, s_0\right)=\frac{1}{8\pi^2} \left(1+\frac{\alpha\left(\sqrt{\tau}\right)}{\pi}\right)
s_0\left(\frac{1}{3}s_0^2+2\tau\right)
+\frac{1}{8\pi^2}\left\langle C_6^v{\cal O}_6^v\right\rangle
\label{m2_vec}
\end{gather}

Consider a two-resonance extension of the narrow resonance model \bref{nar_res} so that \bref{gauss_1res} becomes
\begin{equation}
G^{QCD}\left(\hat s, \tau,s_0\right)=\frac{f_1^2}{\sqrt{4\pi\tau}}\exp{\left[-
\frac{\left(\hat s-m_1^2\right)^2}{4\tau}\right]}
+\frac{f_2^2}{\sqrt{4\pi\tau}}\exp{\left[-
\frac{\left(\hat s-m_2^2\right)^2}{4\tau}\right]}
\label{gauss_2res}
\end{equation}
The corresponding expression for the normalized Gaussian sum-rule then becomes
\begin{equation}
N^{QCD}\left(\hat s, \tau,s_0\right)=\frac{1}{\sqrt{4\pi\tau}}
\left\{
r_1\exp{\left[-
\frac{\left(\hat s-m_1^2\right)^2}{4\tau}\right]}
+r_2\exp{\left[-
\frac{\left(\hat s-m_2^2\right)^2}{4\tau}\right]}
\right\}
\label{norm_gauss_2res}
\end{equation}
where
\begin{equation}
r_1=\frac{f_1^2}{f_1^2+f_2^2}\quad ,\quad r_2=\frac{f_2^2}{f_1^2+f_2^2} \quad ,\quad
r_1+r_2=1\quad .
\label{two_res_strengths}
\end{equation}

If we calculate moments of  \bref{gauss_2res}, we find that in the two-resonance model
\begin{gather}
\frac{M_1}{M_0}=r_1m_1^2+r_2m_2^2=\frac{1}{2}\left(z+ry\right)
\label{centroid}\\
\sigma^2-2\tau=r_1r_2\left(m_1^2-m_2^2\right)^2=\frac{1}{4}y^2\left(1-r^2\right)
\label{2res_width}\\
A_2=r_1r_2\left(r_2-r_1\right)\left(m_1^2-m_2^2\right)^3=-\frac{1}{4}ry^3\left(1-r^2\right)
\label{2res_asymm}
\end{gather}
where
\begin{gather}
r=r_1-r_2
\label{r_def}\\
y=m_1^2-m_2^2
\label{y_def}\\
z=m_1^2+m_2^2\quad .
\label{z_def}
\end{gather}
We see that in the single resonance limit, $\sigma^2=2\tau$ and $A_2=0$, and hence a clear signature of 
the existence of two resonances is $\sigma^2-2\tau\ne 0$.
\footnote{Note that $A_2=0$ for resonances of equal strength $r_1=r_2$, and hence $A_2=0$ does not 
{\em necessarily} imply a single resonance scenario.}   

In Figures \ref{sigma_f0_fig} and \ref{sigma_a0_fig}, $\sigma^2$ 
is plotted as a function of $\tau$ for the optimum values of $s_0$ resulting from  the single-resonance 
$\chi^2$ analyses of
 the $I=0,1$ scalar currents.  
From  Figures \ref{sigma_f0_fig} and \ref{sigma_a0_fig} 
it is evident that $\sigma^2-2\tau$  is significantly different from  zero,
indicating the presence of a second resonance in both the $I=0,1$ scalar channels.  
In  contrast, Figure \ref{sigma_vec_fig}
for the vector channel shows that $\sigma^2$  differs insignificantly  from $2\tau$. Consequently, 
there is  no evidence that a second vector resonance has enough strength  to 
necessitate extension of the phenomenological model, implying that any such states are weak enough to be absorbed 
into the continuum, and providing a validation of the single-resonance analysis.  
Compared to the Laplace sum-rules  which exponentially suppress excited states, 
the potential sensitivity to excited states is a unique feature of the Gaussian sum-rules.

How might we extract predictions of the hadronic parameters
$\{m_1,r_1,m_2,r_2\}$ (or equivalently $\{r,y,z\}$)
from the Gaussian sum-rules \bref{QCD_had}  in a
two narrow resonance model? The key to solving this problem
lies in the behaviour of the peak position
$\hat{s}_{\text{peak}}(\tau,s_0)$ of $N^{\text{QCD}}(\hat{s},\tau,s_0)$.
As noted in Section \ref{sing_res_sec}, in the single narrow resonance model,
the phenomenological side of \bref{norm_gauss} possesses a $\tau$-independent peak
located at $\hat{s}=m_r^2$.
In the double narrow resonance model \bref{norm_gauss_2res}, the situation is more complicated.
For any fixed value of $\tau$ sufficiently large compared to $m_2^2-m_1^2$
(in practice, this is indeed the case), the right-hand-side of \bref{norm_gauss_2res}
exhibits a single peak; however, the position of
this peak varies with $\tau$, and it is this $\tau$-dependent {\em peak drift}
which enables us to extract hadronic parameters in the
double narrow resonance model.

To determine an expression for the phenomenological peak drift, we
find the location of the peak by differentiating the 
right-hand-side of \bref{norm_gauss_2res} and setting the result equal to zero.
In terms of the parameters (\ref{r_def}--\ref{z_def}), this yields
\begin{equation} \label{drifteqn}
  \frac{(r+1)\left(\hat s-\frac{1}{2}z-\frac{1}{2}y\right)}{(r-1)\left(\hat s-\frac{1}{2}z+\frac{1}{2}y\right)}
  - \exp\left[ \frac{y\left(z-2\hat s\right)}{4\tau} \right] = 0 .
\end{equation}
Unfortunately, we cannot explicitly solve~(\ref{drifteqn}) for $\hat{s}$
as a function of $\tau$, so we instead compute the peak drift as a series
expansion in the parameter $1/{\tau}$:
\begin{equation} \label{peakdrift}
  A + \frac{B}{\tau} + \frac{C}{\tau^2},
\end{equation}
where $A$, $B$, and $C$ are functions of the parameters $\{r,y,z\}$, 
and  the quantities $\{B,~C\}$ will be non-zero except in the single resonance limit ($y=0$) 
or in the case of equal resonance strengths ($r=0$).
Explicit numerical testing in worst case scenarios involving two resonances
with masses in the 1--2 GeV range show that the first omitted term
in~(\ref{peakdrift}) [{\it i.e.} \ $D/{\tau^3}$] is negligible provided that
we restrict our attention to $\tau \geq 2\ \text{GeV}^4$.

We define the following $\chi^2$ function which measures quantitatively the
discrepency between the theoretical peak drift contained in
$N^{\text{QCD}}(\hat{s},\tau,s_0)$ and the phenomenological peak drift
approximated by~(\ref{peakdrift})
\begin{equation} \label{driftchi2}
   \chi^2 (s_0) = \sum_{i=1}^N \left[ \hat{s}_{peak} (s_0,\tau_i)
    -A(s_0) -\frac{B(s_0)}{\tau_i} -\frac{C(s_0)}{\tau_i^2} \right]^2,
\end{equation}
where $A(s_0),\ B(s_0),$  and $C(s_0)$ are obtained by solving the
linear system of equations defined by the 
$\chi^2$-minimizing conditions
\begin{equation} \label{conditions}
   \frac{\partial\chi^2}{\partial A} = \frac{\partial\chi^2}{\partial B} =
   \frac{\partial\chi^2}{\partial C} = 0 .
\end{equation}
Minimizing~(\ref{driftchi2}) numerically using the Golden search algorithm \cite{num_recipes}
determines an optimum value for $s_0$.
At this optimum $s_0$, the moments~\bref{mk_gauss} exhibit negligible
dependence on $\tau$ as explicitly occurs in the double resonance model, and thus the moments 
are well approximated on the entire
$\tau$ region of interest by averaged values
\begin{equation} \label{average}
   \overline{M_i(s_0)} = \frac{1}{\tau_f-\tau_i}\int_{\tau_i}^{\tau_f}
    M_i(\tau,s_0) d\tau .
\end{equation}
Next, we calculate $\overline{\sigma^2-2\tau}$ and $\overline{A_2}$
by substituting the average moments~(\ref{average}) into~(79) and~(80).
Finally, we substitute these quantities along with $\overline{M_0}$
and $\overline{M_1}$
into the following inversion formulae which follow easily from (\ref{centroid}--\ref{2res_asymm}):
\begin{gather}
  z  =   2\frac{M_1}{M_0} + \frac{A_2}{\sigma^2-2\tau} 
\label{z_moms}\\
  y  =   \frac{ -\sqrt{A_2^2 + 4(\sigma^2-2\tau)^3}}{\sigma^2-2\tau} 
\label{y_moms}\\
  r  =   \frac{A_2}{\sqrt{A_2^2 + 4(\sigma^2-2\tau)^3}}
\label{r_moms}
\end{gather}
to arrive at our predictions for the hadronic parameters $\{r,y,z\}$
in the double narrow resonance model.

These techniques for analyzing the Gaussian sum-rule  for a two-resonance phenomenological model 
can now be applied to the  $\bar n n$ quark scalar mesons. 
The range $2.0\unt{GeV^4}\le\tau\le4.0\unt{GeV^4}$ will be used to define the $\chi^2$ \bref{driftchi2} 
since it leads to acceptable convergence of the series solution \bref{peakdrift} for the peak drift as well as minimizing
uncertainties from the (large) higher order perturbative coefficients in \bref{gauss_scalar_QCD}.  
Table \ref{scalar_tab} shows the Gaussian sum-rule prediction of the quark scalar resonance
 parameters and continuum  resulting from the
two-resonance $\chi^2$-optimization  procedure.  
Figures \ref{f0_2res_fig} and \ref{a0_2res_fig}  compare the $\hat s$ dependence of the
Gaussian sum-rule and the two-resonance model with the Table \ref{scalar_tab} predictions of the  resonance parameters.
The nearly-perfect agreement  between the theoretical prediction and two-resonance phenomenological model exhibited in 
Figures   \ref{f0_2res_fig} and \ref{a0_2res_fig} is clearly a vast improvement  compared with
 the one-resonance model results displayed in  Figures \ref{f0_sing_res_fig} and \ref{a0_sing_res_fig}, and
provides compelling evidence for the ability of Gaussian sum-rules to predict resonance properties.
The absence of any indication of disagreement between  the theoretical prediction and two-resonance model suggests that any 
further resonances are weak enough to be contained in the QCD continuum.  

The prediction of the resonance parameters for the $\bar n n$ scalar resonances are of phenomenological interest.  The $I=0$ 
states in Table \ref{scalar_tab} suggest that the $f_0(980)$ has a significant $\bar n n$ component, with a weaker excited 
state  with an $\bar n n$ component lying in the vicinity of the $f_0(1370)$ and $f_0(1500)$.  
The absence of a light (substantially less than $1\unt{GeV}$)
$I=0$ state  indicates   a decoupling of a light $\sigma$ from the  
$\bar n n$ scalar currents, suggesting a non-$\bar n n$ interpretation for the $\sigma$.  
The $I=1$ states in Table \ref{scalar_tab} similarly suggest a non-$\bar n n$ interpretation of the $a_0(980)$,
and are consistent with identification of the $a_0(1450)$ as the lightest $I=1$ state with a significant $\bar n n$ 
content.  Our prediction of an $I=1$ state near $1.8\unt{GeV}$ suggests  identification of the 
the $I=1$ state $X(1775)$ \cite{PDG} as a scalar state.  Finally, we note that the Gaussian sum-rule predictions for the 
lightest $I=0,1$ $\bar n n$ scalar mesons confirm the results obtained from  Laplace sum-rules 
\cite{lap_scalar} which, because of the their exponential suppression of excited states, are best suited to 
probing the ground states in these channels.

\begin{table}[hbt]
\begin{tabular}{||c|c|c|c|c|c||}\hline\hline
$I$ & $m_1$ & $m_2$ & $r_1$ & $r_2$ & $s_0$ \\ \hline\hline

$0$ & $0.97\unt{GeV}$ & $1.43\unt{GeV}$ & $0.634$  & $0.366$ &$2.35\unt{GeV^2}$
\\
\hline
$1$ &
$1.44\unt{GeV}$ & $1.81\unt{GeV}$ &  
$0.570$ & $0.430$ & $3.50\unt{GeV^2}$ \\\hline\hline
\end{tabular}
\caption{Results of the two-resonance analysis of the Gaussian sum-rules for the $I=0,1$ 
$\bar n n$ quark scalar mesons resulting from the $\chi^2$-optimization procedure.  
Central values of the QCD parameters have been employed.}
\label{scalar_tab}
\end{table}

As a check on the methods developed for analyzing the two-resonance case, we have also performed a least-squares
fit between the complete two-dimensional $(\hat s,~\tau)$ 
dependence of the two-resonance phenomenological model and the theoretical prediction.
The resonance  parameters and continuum parameters obtained from this analysis differ from those of Table \ref{scalar_tab} 
by at most 10\%, demonstrating that the peak-drift method reproduces a full fitting procedure with minimal 
computational expense, and provides unambiguous results compared to a full multi-dimensional least-squares fit.  
At the very least, the peak-drift method can be used to initialize  a full multi-dimensional least-squares 
fitting procedure.

\section{Conclusions}
\begin{figure}[htb]
\centering
\includegraphics[scale=0.7]{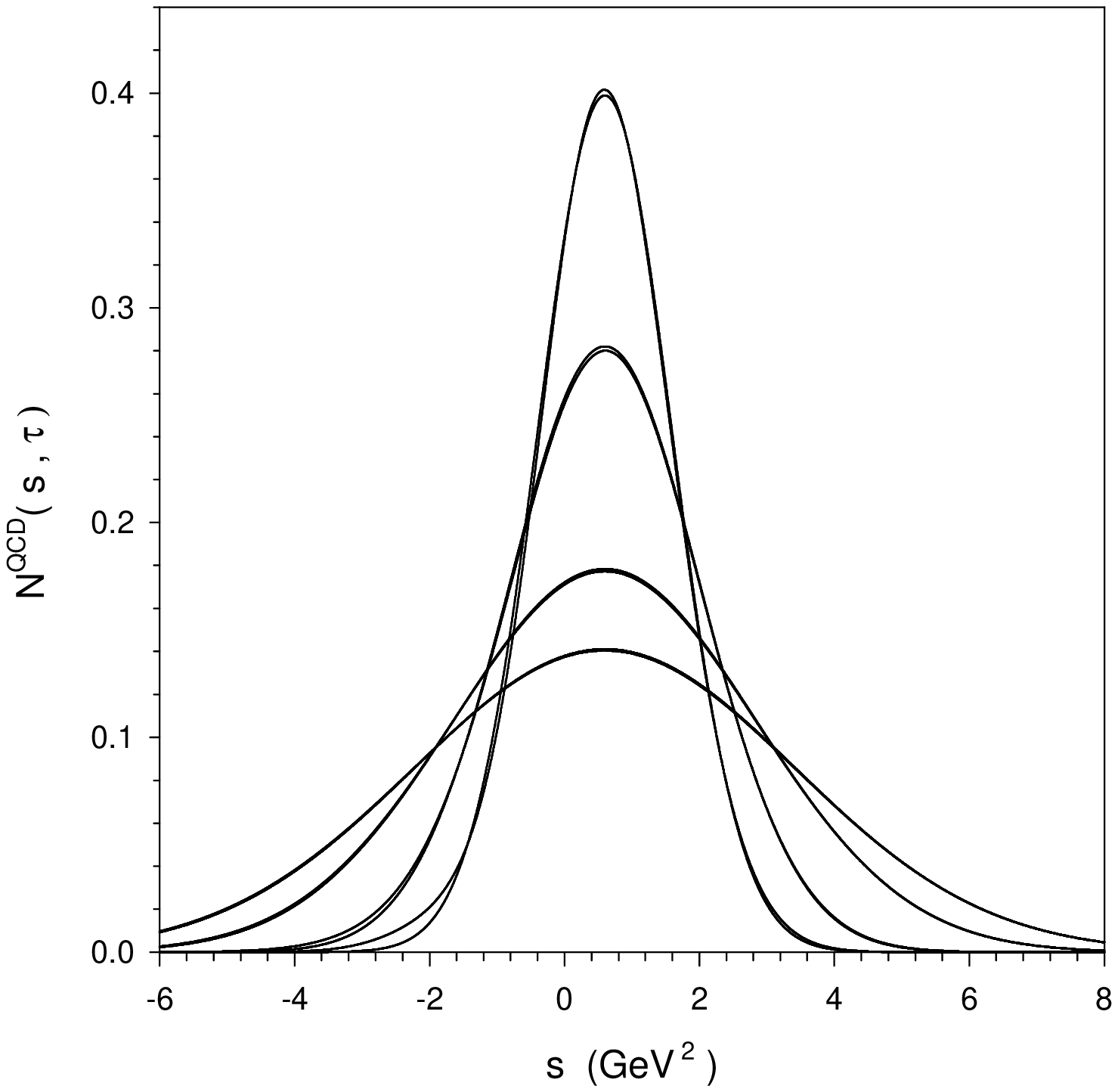}
\caption{Comparison of the vector-current 
theoretical prediction for the $N^{QCD}\left(\hat s, \tau,s_0\right)$ with the
single resonance phenomenological model using  
the $\chi^2$-optimized values of the resonance mass $m_\rho=0.773\unt{GeV}$  
and continuum $s_0=1.22\unt{GeV^2}$.
The $\tau$ values used for the four pairs of curves, from top to bottom in the figure, are respectively
$\tau=0.5\unt{GeV^4}$, $\tau=1.0\unt{GeV^4}$, $\tau=2.5\unt{GeV^4}$, and $\tau=4.0\unt{GeV^4}$.  
Note the almost perfect overlap between the theoretical  prediction and phenomenological model.
Central values of the condensate parameters have been used. 
}
\label{rho_fig}
\end{figure}

\begin{figure}[htb]
\centering
\includegraphics[scale=0.7]{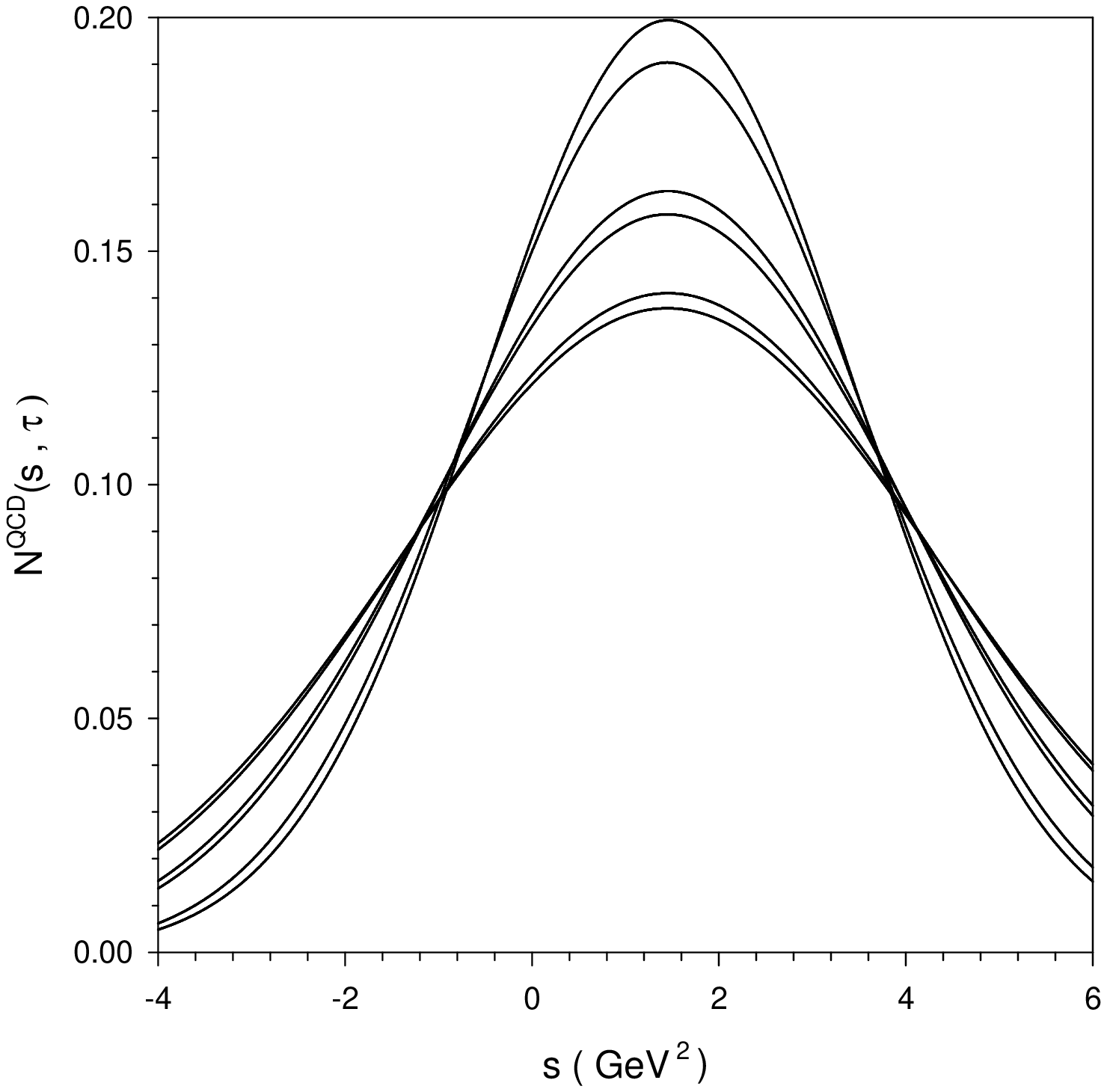}
\caption{Comparison of the $I=0$ scalar current 
theoretical prediction for  $N^{QCD}\left(\hat s, \tau,s_0\right)$ with the
single resonance phenomenological model using  
the $\chi^2$-optimized values of the resonance mass $m=1.21\unt{GeV}$  
and continuum $s_0=2.60\unt{GeV^2}$.
The $\tau$ values used for the three pairs of curves, from top to bottom in the figure, are respectively
 $\tau=2.0\unt{GeV^4}$, $\tau=3.0\unt{GeV^4}$, and $\tau=4.0\unt{GeV^4}$.  
The phenomenological model is consistently {\em larger} than the theoretical prediction near the peak, but is 
consistently {\em smaller} than the 
theoretical prediction in the tails.
Central values of the condensate parameters have been used. 
}
\label{f0_sing_res_fig}
\end{figure}

\begin{figure}[htb]
\centering
\includegraphics[scale=0.7]{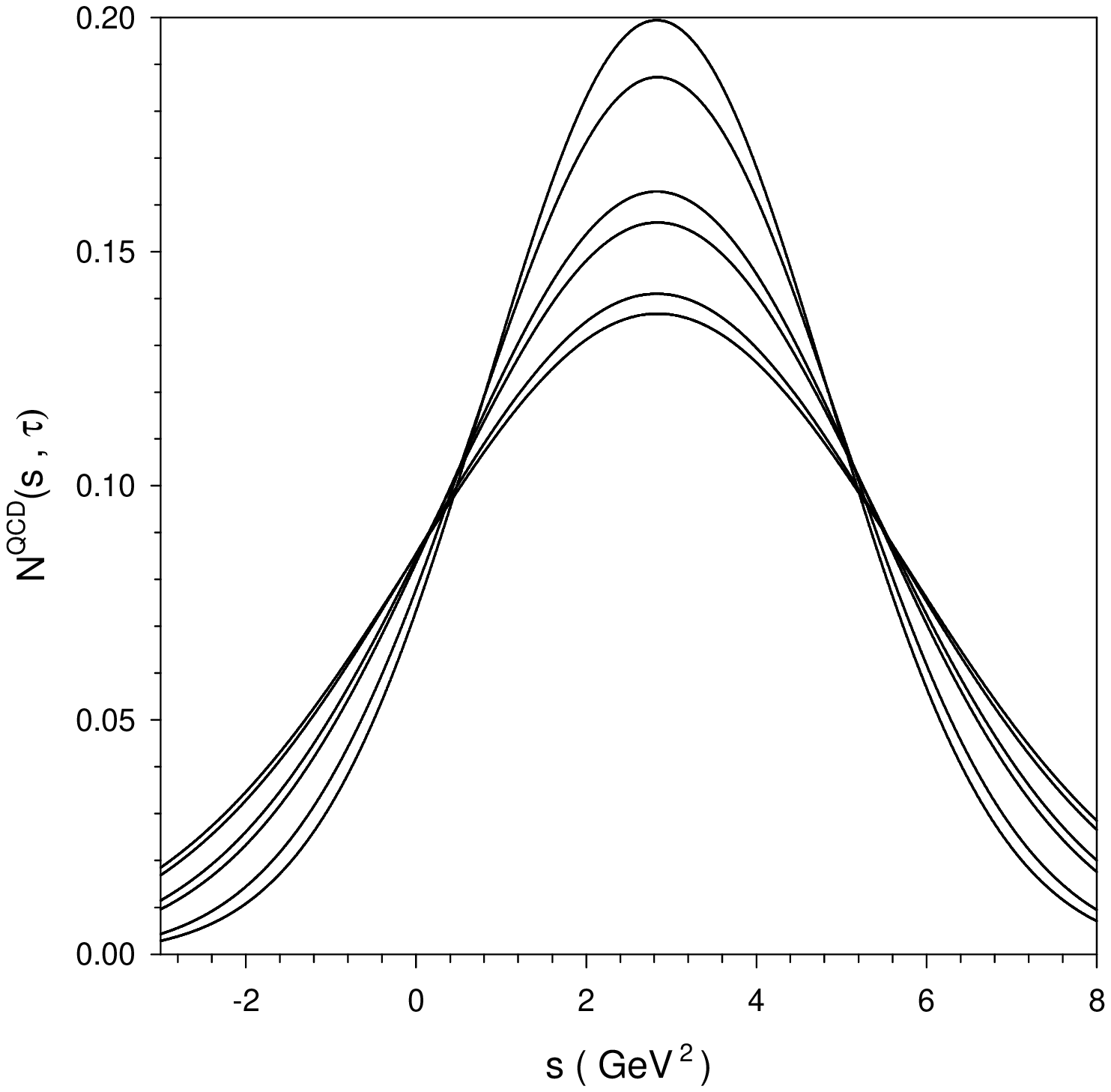}
\caption{Comparison of the $I=1$ scalar current
theoretical prediction for $N^{QCD}\left(\hat s, \tau,s_0\right)$ with the
single resonance phenomenological model using  
the $\chi^2$-optimized values of the resonance mass $m=1.68\unt{GeV}$  
and continuum $s_0=3.9\unt{GeV^2}$.
The $\tau$ values used for the three pairs of curves, from top to bottom in the figure, are respectively
 $\tau=2.0\unt{GeV^4}$, $\tau=3.0\unt{GeV^4}$, and $\tau=4.0\unt{GeV^4}$.  
The phenomenological model is consistently {\em larger} than the theoretical prediction near the peak, but is 
consistently {\em smaller} than the 
theoretical prediction in the tails.
Central values of the condensate parameters have been used. 
}
\label{a0_sing_res_fig}
\end{figure}

In this paper we have developed techniques for using  QCD Gaussian sum-rules to predict hadronic resonance properties, and
discussed the conceptual advantages of the Gaussian sum-rule in providing a natural duality interval for the relation between
QCD and hadronic physics. 
Methods for analyzing a single resonance model based on the $\hat s$ peak position and peak height of the theoretical prediction for the Gaussian sum-rule
were developed and applied to the $\rho$ meson as a test case.  Motivated by the accuracy of the $\rho$ mass prediction and
exceptional agreement between the theoretical prediction and single resonance phenomenological model 
(see Figure~\ref{rho_fig}), the more phenomenologically challenging case of the non-strange $\bar n n$ quark scalar mesons 
was considered.    However, the single resonance analysis of the Gaussian sum-rule of the quark scalar currents 
revealed a discrepancy between the  theoretical prediction and phenomenological model (see Figures \ref{f0_sing_res_fig} and
\ref{a0_sing_res_fig}).  

\begin{figure}[htb]
\centering
\includegraphics[scale=0.7]{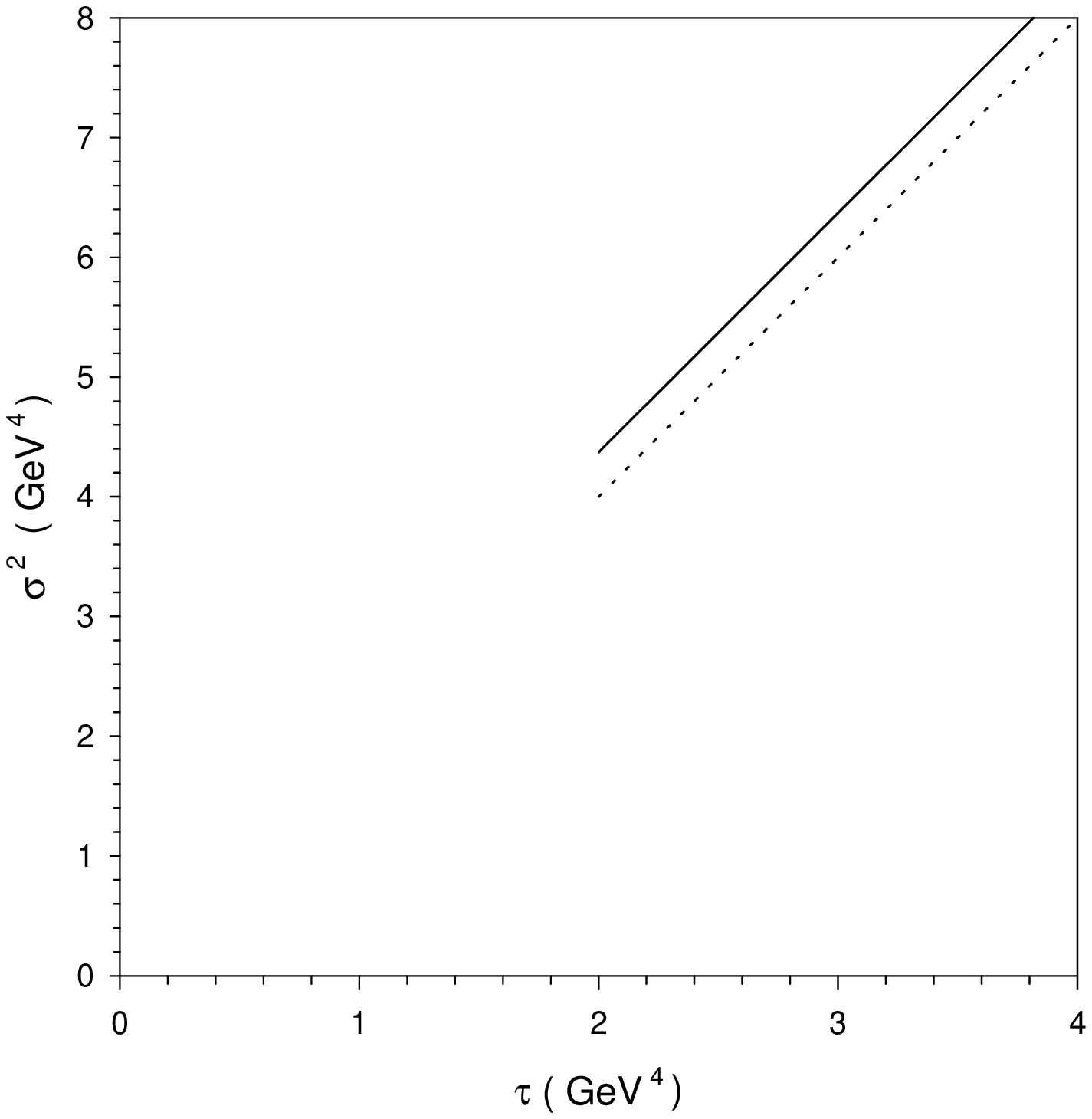}
\caption{Plot of $\sigma^2$ for the theoretical prediction (solid curve) compared with $\sigma^2=2\tau$ 
for the single-resonance model (dashed curve) for the $I=0$ scalar channel
using the $\chi^2$-optimized value of the continuum $s_0=2.6\unt{GeV^2}$.  
}
\label{sigma_f0_fig}
\end{figure}

\begin{figure}[htb]
\centering
\includegraphics[scale=0.7]{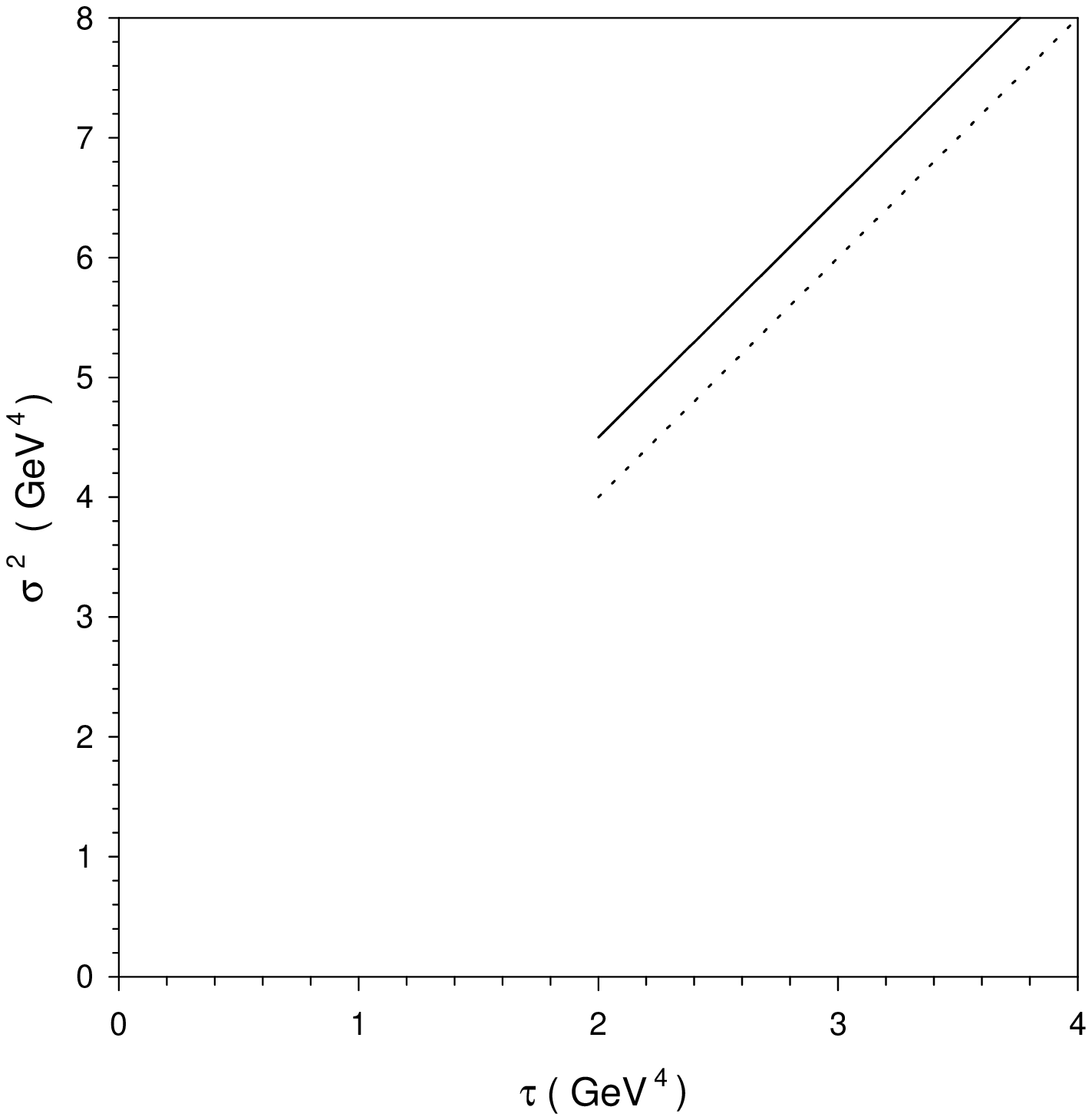}
\caption{Plot of $\sigma^2$ for the theoretical prediction (solid curve) compared with $\sigma^2=2\tau$ 
for the single-resonance model (dashed curve)   for the $I=1$ scalar channel
using the $\chi^2$-optimized value of the continuum $s_0=3.9\unt{GeV^2}$.  
}
\label{sigma_a0_fig}
\end{figure}

\begin{figure}[htb]
\centering
\includegraphics[scale=0.7]{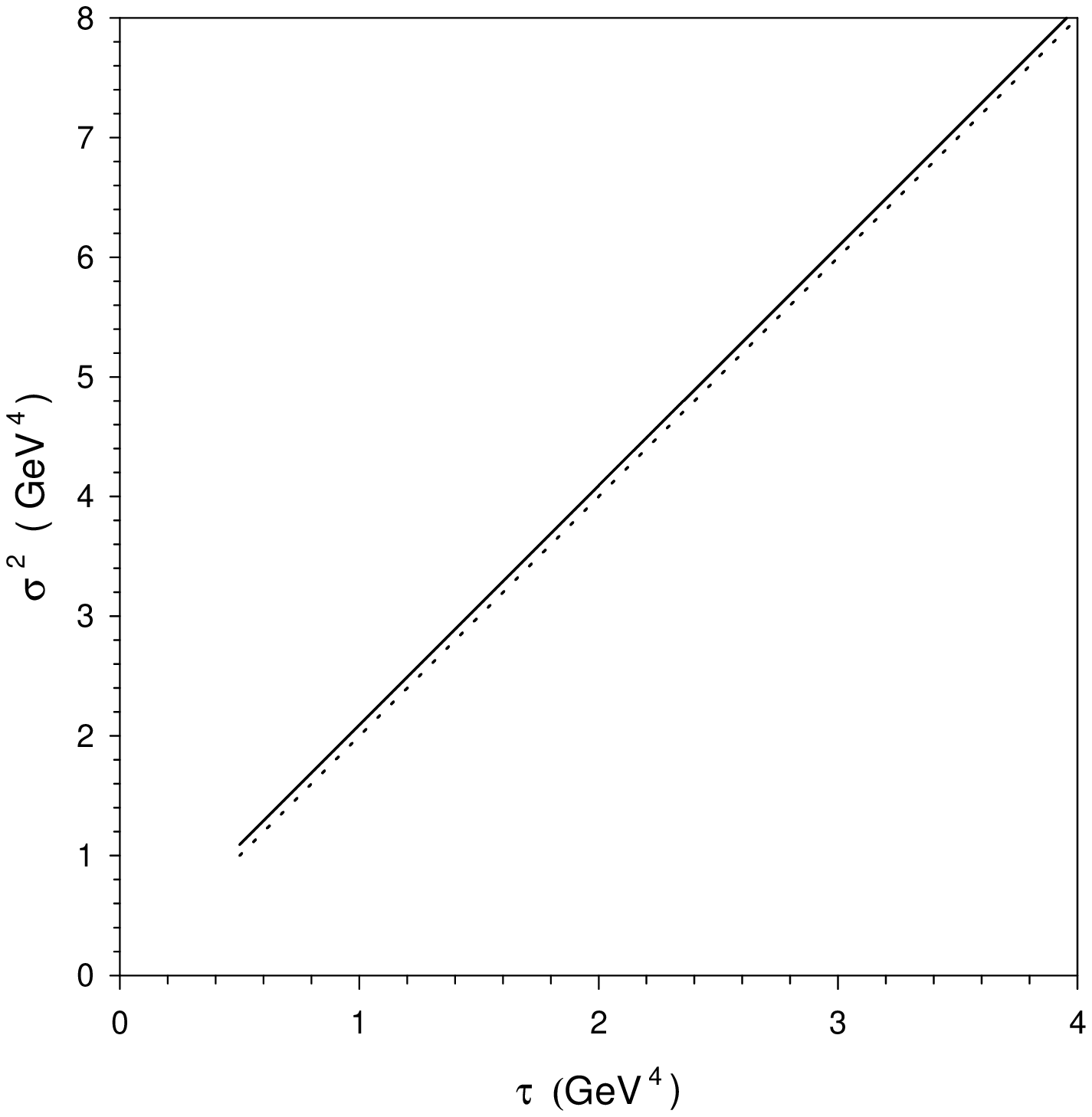}
\caption{
Plot of $\sigma^2$ for the theoretical prediction (solid curve) compared with $\sigma^2=2\tau$ 
for the single-resonance model (dashed curve) for the vector channel
using the $\chi^2$-optimized value of the continuum $s_0=1.22\unt{GeV^2}$.  
}
\label{sigma_vec_fig}
\end{figure}

This discrepancy between the theoretical prediction and single resonance model for the scalar currents 
is addressed by including a second resonance in the phenomenological model. 
The quantity $\sigma^2$, defined in \bref{dist_width} by moments of the 
Gaussian sum-rule, provides a criterion   for evaluating the necessity of  including a second resonance  in the   
phenomenological model.  Figures \ref{sigma_f0_fig}--\ref{sigma_vec_fig} illustrate that the single resonance model is
validated for the vector case, but is insufficient for the scalar currents.  Compared with the Laplace sum-rule, 
the potential sensitivity  of the Gaussian sum-rule to excited states is a novel feature of the Gaussian sum-rules. 

\begin{figure}[htb]
\centering
\includegraphics[scale=0.7]{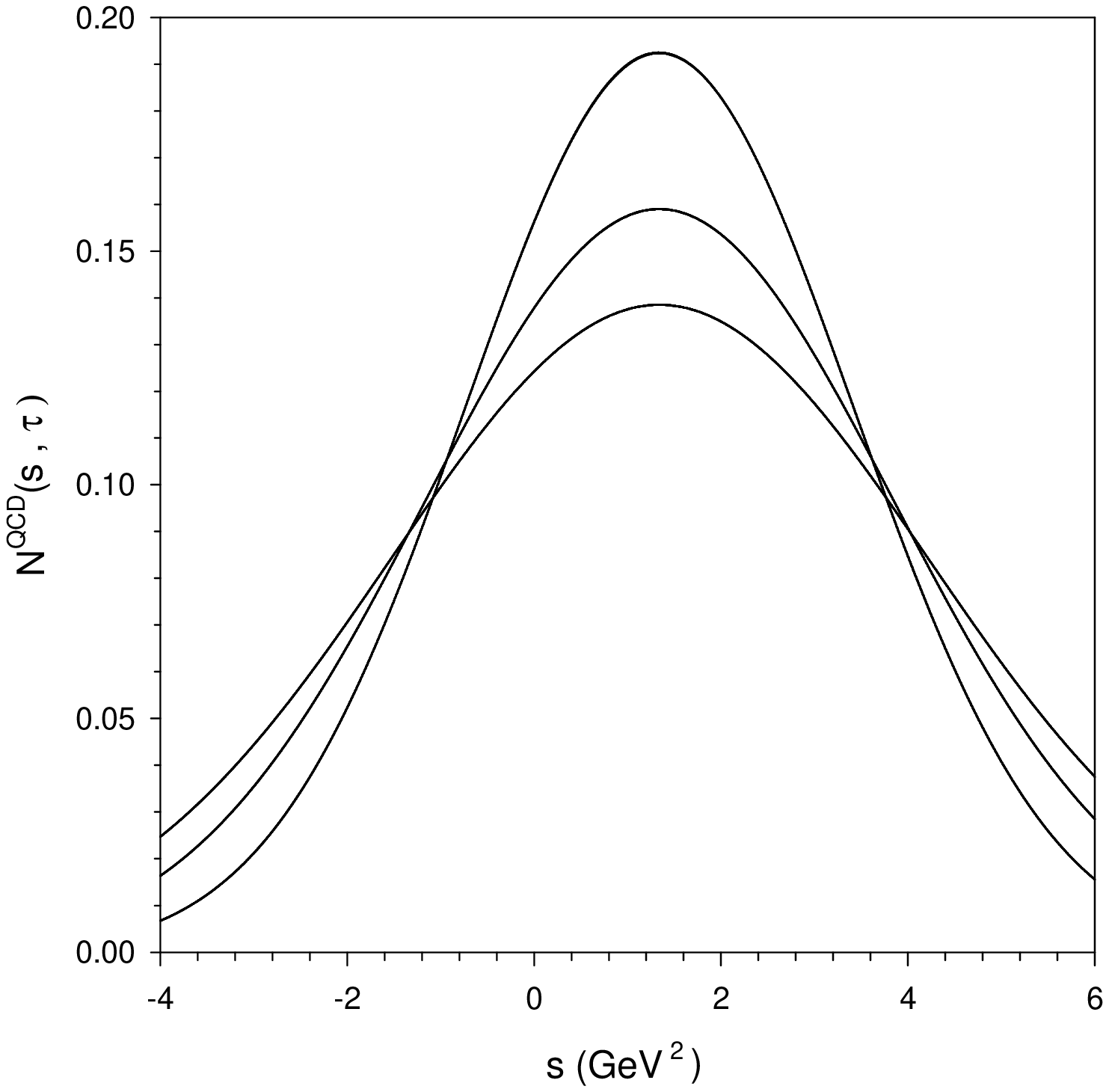}
\caption{Comparison of the $I=0$ scalar-current theoretical prediction $N^{QCD}\left(\hat s, \tau,s_0\right)$ with the
two-resonance phenomenological model using  
the $\chi^2$-optimized values of the resonance masses ($m_1=0.97\unt{GeV}$, $m_2=1.43\unt{GeV}$), 
relative resonance strengths  ($r_1=0.634$, $r_2=0.366$)
and continuum $s_0=2.35\unt{GeV^2}$.
The $\tau$ values used for the four pairs of curves, from top to bottom in the figure, are respectively
 $\tau=2.0\unt{GeV^4}$, $\tau=3.0\unt{GeV^4}$, and $\tau=4.0\unt{GeV^4}$.  
Note the almost perfect overlap between the theoretical  prediction and phenomenological model.
Central values of the condensate parameters have been used. 
}
\label{f0_2res_fig}
\end{figure}

\begin{figure}[htb]
\centering
\includegraphics[scale=0.7]{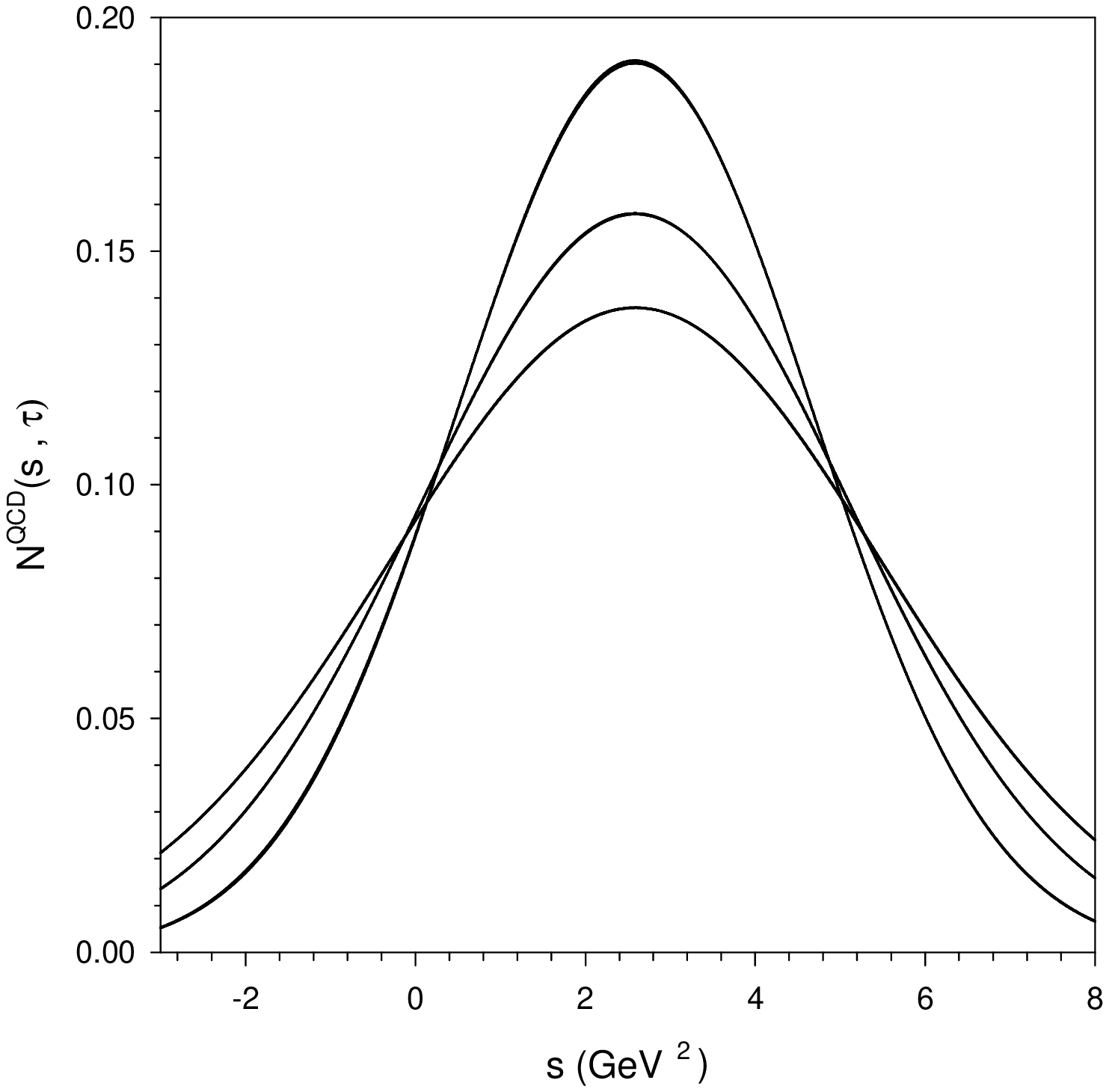}
\caption{Comparison of the $I=1$ scalar current theoretical prediction $N^{QCD}\left(\hat s, \tau,s_0\right)$ with the
two-resonance phenomenological model using  
the $\chi^2$-optimized values of the resonance masses ($m_1=1.44\unt{GeV}$, $m_2=1.81\unt{GeV}$), 
relative resonance strengths  ($r_1=0.570$, $r_2=0.430$)
and continuum $s_0=3.50\unt{GeV^2}$.
The $\tau$ values used for the four pairs of curves, from top to bottom in the figure, are respectively
 $\tau=2.0\unt{GeV^4}$, $\tau=3.0\unt{GeV^4}$, and $\tau=4.0\unt{GeV^4}$.  
Note the almost perfect overlap between the theoretical  prediction and phenomenological model.
Central values of the condensate parameters have been used. 
}
\label{a0_2res_fig}
\end{figure}

The methods developed for analyzing the  Gaussian sum-rules in a single resonance phenomenological model 
can be extended to the two-resonance model by studying the $\tau$ dependence of the $\hat s$ peak position 
(peak drift) of the 
theoretical prediction for the Gaussian sum-rule.  When applied to the Gaussian 
sum-rule of $\bar n n$ quark scalar currents, 
we find the discrepancy between the theoretical prediction and phenomenological model observed in the single-resonance
analysis is resolved by inclusion of a second resonance, and the resulting agreement  
between the  theoretical prediction and phenomenological model (see Figures \ref{f0_2res_fig} and
\ref{a0_2res_fig}) is astounding.  

The phenomenological results for the  $\bar n n$ quark scalar mesons (see Table \ref{scalar_tab}) 
provide valuable information for interpretation of the scalar mesons. In particular, the Gaussian sum-rule analysis is in 
excellent agreement with the identification of the $a_0(1450)$ as the lightest quark scalar meson with a significant
$\bar n n$ content, supporting the conclusions of \cite{lap_scalar,nonet}, and suggests that a light ($\ll 1\unt{GeV}$) 
$I=0$ scalar meson  does not have a significant $\bar n n$ component.  The $I=1$ excited state in Table \ref{scalar_tab}
suggests identification of the $X(1775)$ \cite{PDG} as a scalar state containing   $\bar n n$.

In summary, we conclude that the Gaussian sum-rules are not only  useful in their well-established connection with
the finite-energy sum-rule constraint on the continuum threshold \cite{gauss}, but are a valuable technique for studying 
hadronic physics, and should be considered as a viable and complementary alternative to Laplace sum-rules.

\smallskip
\noindent
{\bf Acknowledgements:}
TGS is grateful for  the warm hospitality of GO and W.\ Leidemann  at the   
Universit\`a di Trento while this research was 
conducted.  
TGS and DH  also acknowledge research support 
from the Natural Sciences and Engineering Research Council of Canada (NSERC).

\clearpage

\end{document}